\pdfoutput=1

\documentclass[%
 reprint,
superscriptaddress,
amsmath,amssymb,amsbsy,aps,
 prb,
]{revtex4-1}

\usepackage[pdftex]{graphicx}
\usepackage{epstopdf}
\usepackage{rotating}
\usepackage{dcolumn}
\usepackage{bm}
\usepackage{mathrsfs}
\usepackage{color}
\usepackage{leftidx}
\usepackage{bbm}
\usepackage{verbatim}
\usepackage{soul,xcolor}

\usepackage[caption=false]{subfig}

\usepackage{fixltx2e}
\definecolor{greens}{rgb}{0,0.7,0}

\usepackage[breaklinks,colorlinks=true,linkcolor=blue,urlcolor=blue,citecolor=blue]{hyperref}

\begin{document}


\title{Theory of Large Intrinsic Spin Hall Effect in Iridate Semimetals}

\author{Adarsh S. Patri}
\affiliation{Department of Physics and Centre for Quantum Materials, University of Toronto, Toronto, Ontario M5S 1A7, Canada}

\author{Kyusung Hwang}
\affiliation{Department of Physics and Centre for Quantum Materials, University of Toronto, Toronto, Ontario M5S 1A7, Canada}
\affiliation{Department of Physics, The Ohio State University, Columbus, OH 43210, USA}

\author{Hyun-Woo Lee}
\affiliation{Department of Physics and Centre for Quantum Materials, University of Toronto, Toronto, Ontario M5S 1A7, Canada}
\affiliation{PCTP and Department of Physics, Pohang University of Science and Technology, Pohang 37673, Korea}

\author{Yong Baek Kim}
\affiliation{Department of Physics and Centre for Quantum Materials, University of Toronto, Toronto, Ontario M5S 1A7, Canada}

\date{\today}

\begin{abstract}
We theoretically investigate the mechanism to generate large intrinsic spin Hall effect in iridates or more broadly in 5d transition metal oxides with strong spin-orbit coupling. We demonstrate such a possibility by taking the example of orthorhombic perovskite iridate with nonsymmorphic lattice symmetry, SrIrO$_3$, which is a three-dimensional semimetal with nodal line spectrum. It is shown that large intrinsic spin Hall effect arises in this system via the spin-Berry curvature originating from the nearly degenerate electronic spectra surrounding the nodal line. This effect exists even when the nodal line is gently gapped out, due to the persistent nearly degenerate electronic structure, {suggesting a distinct robustness.} The magnitude of the spin Hall conductivity is shown to be comparable to the best known example such as doped topological insulators and the biggest in any transition metal oxides. To gain further insight, we compute the intrinsic spin Hall conductivity in both of the bulk and thin film systems. We find that the geometric confinement in thin films leads to significant modifications of the electronic states, leading to even bigger spin Hall conductivity in certain cases. We compare our findings with the recent experimental report on the discovery of large spin Hall effect in SrIrO$_3$ thin films.
\end{abstract}

\maketitle

\section{Introduction}

Relativistic spin-orbit coupling (SOC), once relegated to the annals of atomic physics, has become prevalent in modern condensed matter physics. The spin Hall effect,\cite{Sinova2015} as well as its once-contentious cousin the anomalous Hall effect,\cite{Nagaosa2010} are examples of phenomena that are deeply rooted in the physics of SOC. In the spin Hall effect (SHE), an unpolarized electron charge current passing through a material which possesses strong SOC gives rise to a purely spin polarized current in the transverse direction; the inverse spin Hall effect (ISHE) is, as the name suggests, the inverse process where a pure spin current gives rise to a transverse charge current.\cite{Sinova2015,Hirsch_1999,Sinova2004,spintronics2004,spindevices2012}

The SHE can be broadly classified under two regimes: intrinsic (which is dependent solely on band topology and SOC) and extrinsic (where impurity scattering is key).\cite{Sinova2015,she_intrinsic} A common figure of merit used to examine the efficiency of the spin/charge current conversion is the spin Hall angle: $\theta_{\textup{SH}}=\sigma_{\textup{SH}}/{\sigma}$, where $\sigma_{\textup{SH}}$ and $\sigma$ are the respective spin-Hall and charge conductivities.\cite{she_angle} 
{Large SHE has primarily been predicted and/or observed in heavy elemental metals (that possess large SOC) such as Pt,\cite{spintronics_2008, spintronics_2009, Pt_SHE_2007, Guo2008, Pt_SHE_2011, Pt_RB_2011, Pt_RB_2012} Au \cite{spintronics_2008,spintronics_2009,Au_yao_theory} (although the origin of the observed large SHE in Au \cite{Au_SHE_2008} is still under debate \cite{Au_SHE_kondo, Au_SHE_Seki_correction, Au_SHE_japan_correction}), and Pd.\cite{spintronics_2008,spintronics_2009, Pd_SHE_2010} More recently, there have also been suggestions of large SHE in topological insulators from spin-torque measurements in topological insulator Bi$_2$Se$_3$ \cite{SHE_BiSe_2014} and Cr-doped Bi$_2$Se$_3$ thin films.\cite{SHE_doped_BiSe_2014}
However, despite these two classes of materials demonstrating such large SHEs, they do have their respective drawbacks: the large charge conductivity of the heavy metals hinders their efficiency as spin current detectors (since the spin Hall resistivity, a figure of merit for spin detection, is proportional \cite{Fujiwara_2013} to $\sigma_{\textup{SH}}/\sigma^2$)},
and the coupling of magnetic layers to topological insulator thin films once again forms a bottleneck for the spin current generation efficiency.\cite{SHE_BiSe_2014} In an attempt to achieve a large spin Hall angle with better efficiency, a new class of 5d transition metal oxides has recently been investigated.\cite{Fujiwara_2013} These materials not only have large SOC (due to the 5d nature of the conduction bands) but are also useful due to their tunability of their physical properties via their magnetic ordering and/or correlation effects.\cite{XW_2011,Nd2Ir2O7_Ueda_2015, Nd2Ir2O7_Tian_2016} Moreover, the emergent semimetallic nature of numerous types of iridates \cite{Carter_2012, Chen_2015, Optical_weyl_2015, Ueda_2012, Nd2Ir2O7_Ueda_2015, Nd2Ir2O7_Tian_2016}  provides hope that the spin Hall angle for the iridates will be superior to that of the heavy elemental metals, due to semimetals' relatively smaller charge conductivity.

\begin{figure*}
\centering
\includegraphics[width=\linewidth]{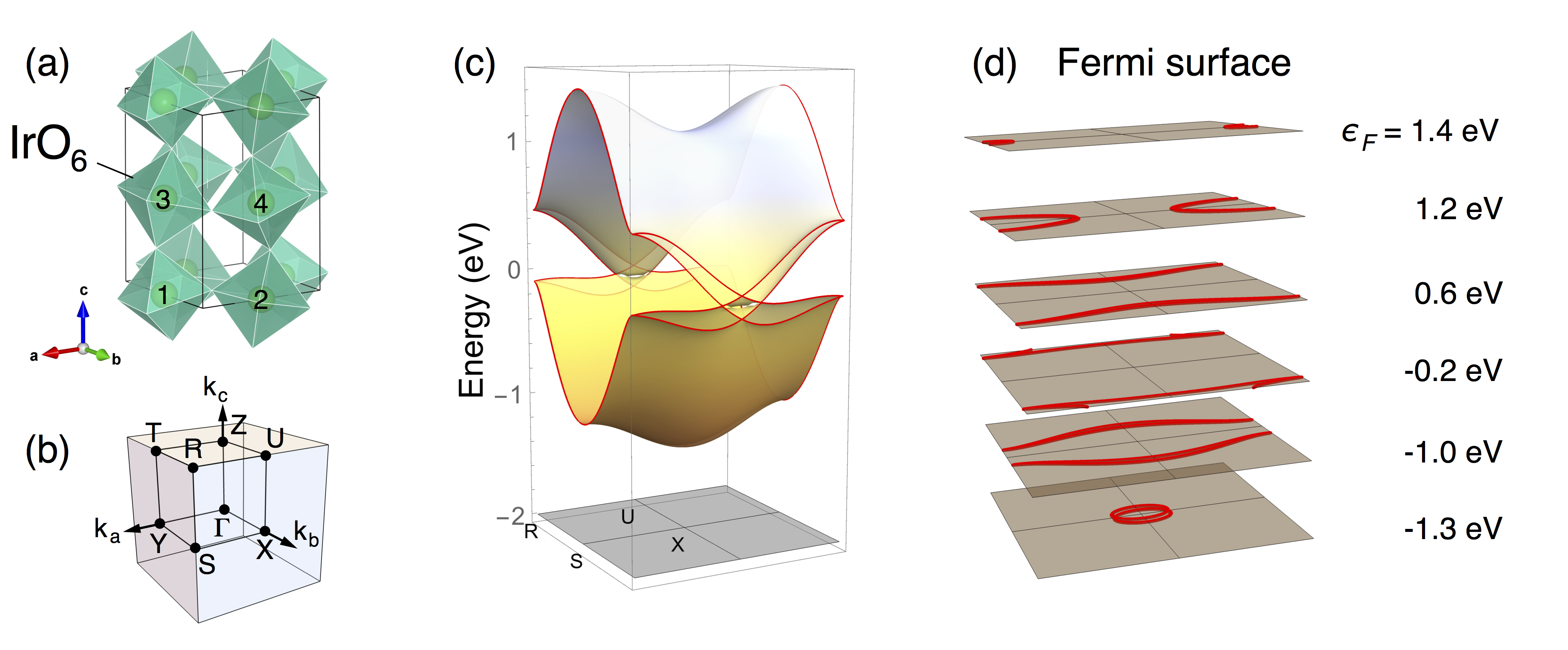}
\caption{Crystal structure and electron energy bands of orthorhombic perovskite SrIrO$_3$.
\textcolor{black}{(a) Orthorhombic unit cell with four Ir sublattices (numbered) and oxygen octahedra surrounding the Ir sites. (b) Brillouin zone, with labelled points Y=($\pi$,0,0), X=(0,$\pi$,0), Z=(0,0,$\pi$), R=($\pi$,$\pi$,$\pi$), U=(0, $\pi$,$\pi$) in the coordinate of $(k_a,k_b,k_c)$ [=$({\bf k}\cdot{\bf a}, {\bf k}\cdot{\bf b}, {\bf k}\cdot{\bf c})$ where  $\{{\bf a},{\bf b},{\bf c}\}$ are the orthorhombic lattice vectors]. (c) Electron energy bands at the $k_b=\pi$ plane (R-U-X-S) of the Brillouin zone. (d) Fermi surface cross sections \textcolor{black}{within} the $k_b=\pi$ plane for various Fermi energies.}} 
\label{fig:structure}
\end{figure*}

In this work, we theoretically investigate the intrinsic spin Hall effect in orthorhombic perovskite SrIrO$_3$, which is a 5d transition metal oxide with strong SOC. Our work is motivated by recent experimental work on SrIrO$_3$ thin films, where one of the largest ever recorded spin Hall conductivity was measured.\cite{SIO_SHE_exp} Interestingly, SrIrO$_3$ possesses a topological nodal line (a loop of band crossings) that is protected by nonsymmorphic symmetries.\cite{Carter_2012, Chen_2015, Chen_2016, Fang_2015, Rhim_YB_2015, Rhim_YB_2016, Huh_2016} As shown later, the persistent nearly degenerate electronic structure near the nodal line plays an important role in the SHE in SrIrO$_3$.
We study the SHE in the framework of linear response theory, focusing on both the bulk system as well as thin film configurations to enable comparison with experiment. In both cases, we compute the spin Hall conductivity by utilizing the $j_{\textup{eff}}=1/2$ tight binding model constructed from \textit{ab initio} studies of SrIrO$_3$.\cite{Carter_2012,Chen_2015} For the bulk system, we also consider the impact of breaking various nonsymmorphic symmetries of the \textit{Pbnm} space group on the spin Hall conductivity. In this way, we are able to study the robustness of the spin Hall conductivity to a variety of symmetry breaking perturbations. Finally, we consider (010)$_c$ thin film of SrIrO$_3$, where pseudo-cubic coordinate system is used to describe the thin film direction, which enables a more direct connection to the recent experimental work. 

Our computations predict the spin Hall conductivity in the bulk system to be remarkably large [$\sigma_{\textup{SH}}^{\textup{bulk}}\sim10^4(\hbar/e)(\Omega\textup{m})^{-1}$], \textcolor{black}{comparable to the spin Hall conductivity in the heavy elemental metal family}. Moreover, we demonstrate that the bulk spin Hall conductivity is robust and stable despite the introduction of various symmetry breaking terms, as well as lifting of the gapless nodal line, due to the persisting nearly degenerate electronic spectra. Our thin film calculations predict a large film spin Hall conductivity [$\sigma_{\textup{SH}}^{\textup{film}}\sim10^4(\hbar/e)(\Omega\textup{m})^{-1}$] in the configuration corresponding to the experiment, which is at least one order of magnitude greater than the same configuration in the bulk system. We attribute this surprising enhancement to the significant modification of the bulk-like eigenstates in the film (due to the restricted geometry breaking certain lattice symmetries). The enormity of the spin Hall conductivity predicted in this study for both the bulk and thin film systems, 
as well as the robustness of their response, promises a bright and exciting future for the family of 5d transition metal oxides in the field of spintronics. 

The rest of the paper is organized as follows. In Sec. \ref{sec:model_H}, we describe the important properties of the tight binding Hamiltonian (including the $Pbnm$ space group that it belongs to), as well as the key features in the band structure. In Sec. \ref{shc_bulk}, we present the Kubo formula used to compute the spin Hall conductivity, and the spin Hall conductivity as a function of the Fermi level for different directions. We also elucidate on which regions contribute the most to the spin Hall conductivity, to explain the unexpectedly large values, and discuss the role of the nodal line in the spin Hall conductivity. In Sec. \ref{shc_bulk_break}, we incorporate various types of symmetry breaking terms into our bulk model calculation, and study the reasons for the small (less than order of magnitude) change despite the introduction of these terms. In Sec. \ref{shc_film}, we examine the spin Hall conductivity in the (010)$_c$ thin film configuration. Lastly, in Sec. \ref{conclusion}, we \textcolor{black}{summarize our results and discuss their relevance}, as well as provide direction for future work.

\section{Model Hamiltonian} \label{sec:model_H}

\begin{table*}[t]
\caption{
{
$Pbnm$ space group symmetry and remaining symmetries in various symmetry-broken systems. In the second column, the symmetry operations are defined by the transformation rules of {the} position vector (${\bf R} = a \hat{a} + b \hat{b} + c \hat{c} \rightarrow {\bf R'} = a' \hat{a} + b' \hat{b} + c' \hat{c}$).
The last four columns represent the remaining symmetries in {the} symmetry-broken bulk systems (Sec. \ref{shc_bulk_break}) and (010)$_c$ thin film (Sec. \ref{shc_film}) with checkmarks.}}
\begin{ruledtabular}
\begin{tabular}{cccccc}
Symmetry & ${\bf R'}$ & $H + h_{gap}$ [Bulk] & $H + h_{xz}$ [Bulk] & $H + h_{xx}$ [Bulk] & (010)$_c$ [Film] 
\\
\hline
$n$-glide ($G_n$) & $a+\frac{1}{2}$, $- b+\frac{1}{2}$, $c+\frac{1}{2}$ & & & & 
\\ 
$b$-glide ($G_b$) & $-a+\frac{1}{2}$, $b+\frac{1}{2}$, $c$ & & \checkmark & \checkmark & 
\\
Mirror ($m$) & $a$, $b$, $-c+\frac{1}{2}$ & \checkmark & & & \checkmark 
\\
Inversion ($\bar{I}$) & $-{a}$, $-{b}$, $-{c}$  &  \checkmark & \checkmark & \checkmark & \checkmark
\\
$a$-screw ($S_a$) & $a+\frac{1}{2}$, $-b+\frac{1}{2}$, $-{c}$ & & \checkmark & \checkmark &
\\
$b$-screw ($S_b$) & $-a+\frac{1}{2}$, $b+\frac{1}{2}$, $-c+\frac{1}{2}$ & & & &  
\\
$c$-screw ($S_c$) & $-{a}$, $-{b}$, $c+\frac{1}{2}$ & \checkmark & & & \checkmark
\\
\end{tabular}
\end{ruledtabular}
\label{tab:Pbmn}
\end{table*}

We employ the tight-binding model constructed in Refs. \onlinecite{Carter_2012,Chen_2015} to describe the electronic structure of SrIrO$_3$.
Due to the significant tilting and rotation of oxygen octahedra, the system has the orthorhombic perovskite crystal structure with four Ir sublattices and $Pbnm$ nonsymmorphic space group (Fig. \ref{fig:structure}). In the basis of the $j_{\textup{eff}}=1/2$ states for Ir$^{4+}$ electrons,
the model incorporates various electron hopping channels allowed in the orthorhombic perovskite SrIrO$_3$ with the following form of Hamiltonian.
\begin{equation}
H = \sum_{\bf k} \psi_{\bf k}^{\dagger} H_{\bf k} \psi_{\bf k}.
\label{eq:model}
\end{equation}
Here, $\psi=(\psi_{1\uparrow},\psi_{2\uparrow},\psi_{3\uparrow},\psi_{4\uparrow},\psi_{1\downarrow},\psi_{2\downarrow},\psi_{3\downarrow},\psi_{4\downarrow})^T$ are electron operators with the subscripts referring to the Ir sublattice (1,2,3,4) and $j_{\textup{eff}}=1/2$ pseudo-spin ($\uparrow,\downarrow$), and {\bf k} is crystal momentum.
The matrix $H_{\bf k}$ contains ten different hopping channels up to the next nearest neighbor.
Depending on whether the pseudo-spin changes during hopping processes or not, the hopping channels are classified into spin-dependent hopping $\{ t'_p, t_{1p}^o, t_{2p}^o, t_{z}^o, t_{d}^o \}$ and spin-independent hopping $\{t_p, t_z, t_{xy}, t_d, t'_d\}$.
The oxygen octahedron tilting and rotation generate the spin-dependent hopping which is crucial for the SHE in SrIrO$_3$.
Such spin-dependent hopping is not allowed in the perfect cubic perovskite.
For the hopping parameters, we use the values obtained in Ref. \onlinecite{Carter_2012} based on {\it ab initio} calculations.
The explicit form of $H_{\bf k}$ and values of the hopping parameters are presented in Appendix \ref{app_H}. 

The $Pbnm$ space group symmetry, discussed above, dictates the relations between the components of the spin Hall conductivity tensor, as well as protects various features of the band structure. 
{To aid in the discussion to follow in the upcoming sections, we present in Table \ref{tab:Pbmn} a summary of the $Pbnm$ space group{,} as well as the remaining symmetries in various symmetry-broken bulk systems (Sec. \ref{shc_bulk_break}) and thin film (Sec. \ref{shc_film}).}

Figure \ref{fig:structure} depicts the electron band structure of the system at the particular $k_b=\pi$ plane (R-U-X-S) in the Brillouin zone.
There are four doubly degenerate bands on account of four sublattices and Kramers degeneracy. One remarkable feature is the band crossing occurring along a ring about the U point.\cite{Carter_2012,Chen_2015} This ``nodal ring'' is quite small, but protected by nonsymmorphic symmetries compatible with the \textcolor{black}{$k_b= \pi$} plane.\cite{Chen_2016}  
Another interesting point is the ``near-degeneracy" of the four bands found at the $k_b=\pi$ plane and $k_a=\pi$ plane (T-R-S-Y) as highlighted by cyan in Fig. \ref{fig:band-SHC}. It must be noted both features appear by spin-dependent electron hopping in the system.  \textcolor{black}{Interestingly, the nearly degenerate structure and the size of the nodal ring are correlated and controlled by the same hopping parameter $t_d^o$. In fact, this intimate correlation between the two features prevails in a wide window of permissible values of $t_d^o$, so that the ring appears concurrently with the presence of the nearly degenerate structure (provided the necessary nodal ring symmetry is intact). Even when the ring is gently gapped out (as will be seen) the nearly degenerate structure is preserved.} Such a correlated structure, however, will eventually be suppressed by a large perturbation which may significantly modify the underlying band structure.

\subsection{Symmetry-Protected Nodal Line}

Before we move on to the next section, we briefly review the mechanism of the symmetry-protected nodal line \cite{Chen_2016} for self-containedness in our discussions.
The nodal line band crossing {(or nodal ring)} occurs by the interplay of the three nonsymmorphic symmetries, $n$-glide plane ($G_n$), $b$-glide plane ($G_b$), and $a$-screw axis ($S_a$), and its symmetry protection can be understood by investigating the little group of the Hamiltonian matrix $H_{\bf k}$.

\begin{table}[h]
\centering
\caption{Elements of the little group of $H_{\bf k}$ at various high symmetry locations in the Brillouin zone. In each case, little group elements are denoted with checkmarks.}
\begin{ruledtabular}
\begin{tabular}{ccccc}
& {\bf k} & $G_n$ & $G_b$ & $S_a$ 
\\
\hline
$k_b=\pi$ plane & $(k_a,\pi,k_c)$ & \checkmark &  &
\\
U point & $(0,\pi,\pi)$ & \checkmark & \checkmark & \checkmark
\\
RS line & $(\pi,\pi,k_c)$ & \checkmark & \checkmark &
\\
SX line & $(k_a,\pi,0)$ & \checkmark &  & \checkmark \\
\end{tabular}
\end{ruledtabular}
\label{tab:littlegroup}
\end{table}
\begin{figure}[b]
\centering
\includegraphics[width=0.92\linewidth]{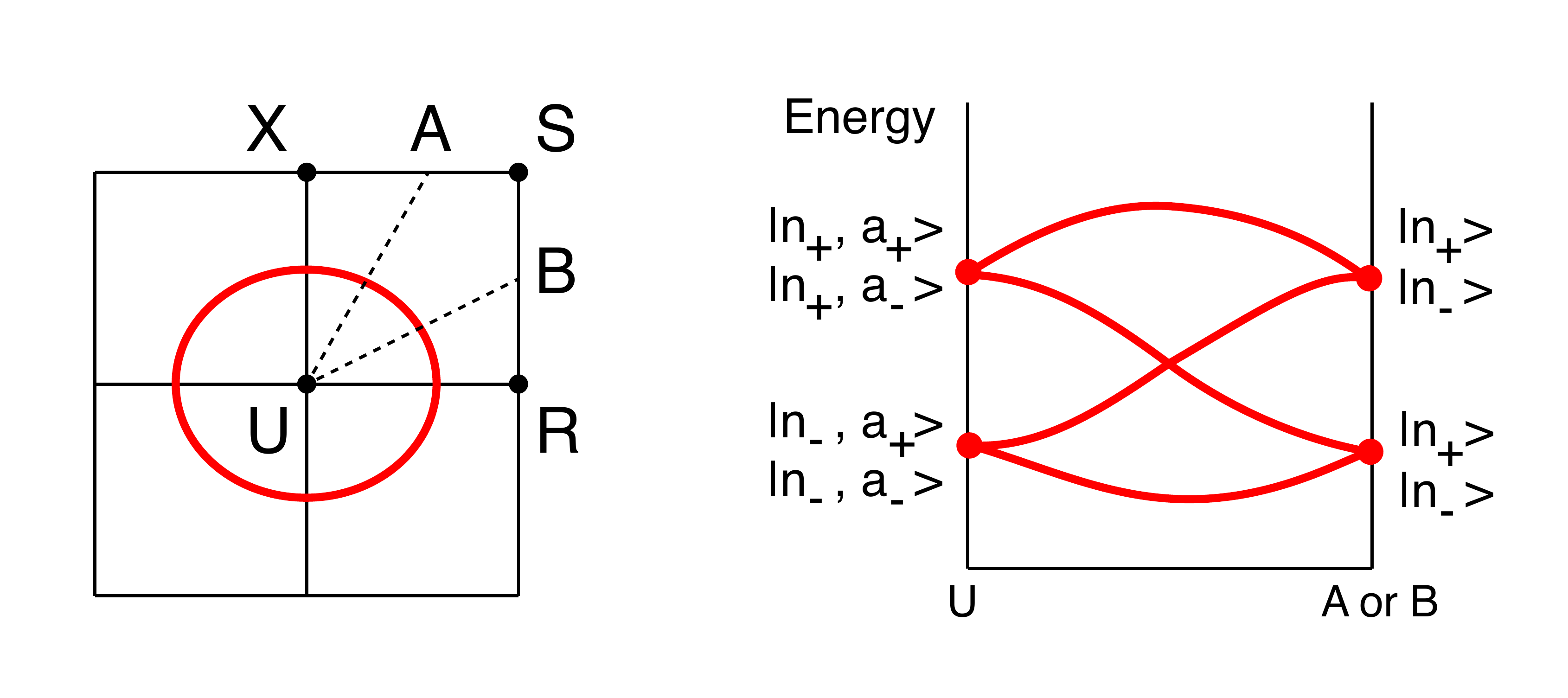}
\caption{Schematic illustration of the mechanism of the symmetry-protected nodal line. In the left figure, the red circle represents the nodal ring on the $k_b=\pi$ plane. The right figure describes how the $G_n$-eigenvalue structure of the energy bands changes along paths connecting the U point with the RS or SX line (dashed lines in the left figure).}
\label{fig:nodalline}
\end{figure}

First, we specify the little group for the $k_b=\pi$ plane where the nodal line appears.
Under a space group symmetry operation, Bloch states (momentum eigenstates) generally move from a ${\bf k}$ point to another in the Brillouin zone unless the operation is a pure translation.
Nevertheless, at high symmetry points of the Brillouin zone, the momentum of Bloch states can be invariant under certain symmetry operations.
Such symmetry operations define the little group of $H_{\bf k}$ at a given high symmetry point.
In the case of the $Pbnm$ space group, the entire $k_b=\pi$ plane is invariant under $G_n$.
Moreover, high symmetry points on that plane such as the U point and RS and SX lines have further little group elements ($G_b$ and/or $S_a$) as summarized in Table \ref{tab:littlegroup}.
One can check this from the transformation rules of electron operators: 
\begin{equation}
\begin{array}{ccl}
G_n: \psi_{\bf k} &\rightarrow& \frac{i}{\sqrt{2}}e^{i\frac{k_a-k_b+k_c}{2}} (\sigma_x-\sigma_y) \nu_x \tau_x \psi_{(k_a,-k_b,k_c)},
\\
G_b: \psi_{\bf k} &\rightarrow& -\frac{i}{\sqrt{2}}e^{i\frac{-k_a+k_b}{2}} (\sigma_x+\sigma_y) \tau_x \psi_{(-k_a,k_b,k_c)},
\\
S_a: \psi_{\bf k} &\rightarrow& -\frac{i}{\sqrt{2}}e^{i\frac{k_a-k_b}{2}} (\sigma_x+\sigma_y) \tau_x \psi_{(k_a,-k_b,-k_c)},
\end{array}
\label{eq:symm-transformation}
\end{equation}
{where $\sigma,~\mu,~\tau$ represent Pauli matrices acting on spin and sublattice degrees of freedom (see Appendix \ref{app_H} for the definitions).}
Using these transformation rules, one can show that $G_n^2= -T_{{\bf a}+{\bf c}}$, $G_b^2=-T_{{\bf b}}$, $S_a^2=-T_{{\bf a}}$ (where $T_{\bf r}$ represents a translation by a lattice vector ${\bf r}$, and the minus sign in each case arises due to a $2\pi$-rotation of $j_{\textup{eff}}=1/2$ spin). 
This tells us the eigenvalues of $\{G_n,G_b,S_a\}$: $n_{\pm}\equiv\pm i e^{i(k_a+k_c)/2}$ for $G_n$, $b_{\pm}\equiv\pm i e^{ik_b/2}$ for $G_b$, and $a_{\pm}\equiv\pm i e^{ik_a/2}$ for $S_a$.
Here it is important to notice that $G_n(=n_{\pm})$ serves as a good quantum number to specify Bloch states over the whole $k_b=\pi$ plane.

\begin{figure*}[t]
\centering
\includegraphics[width=0.8\linewidth]{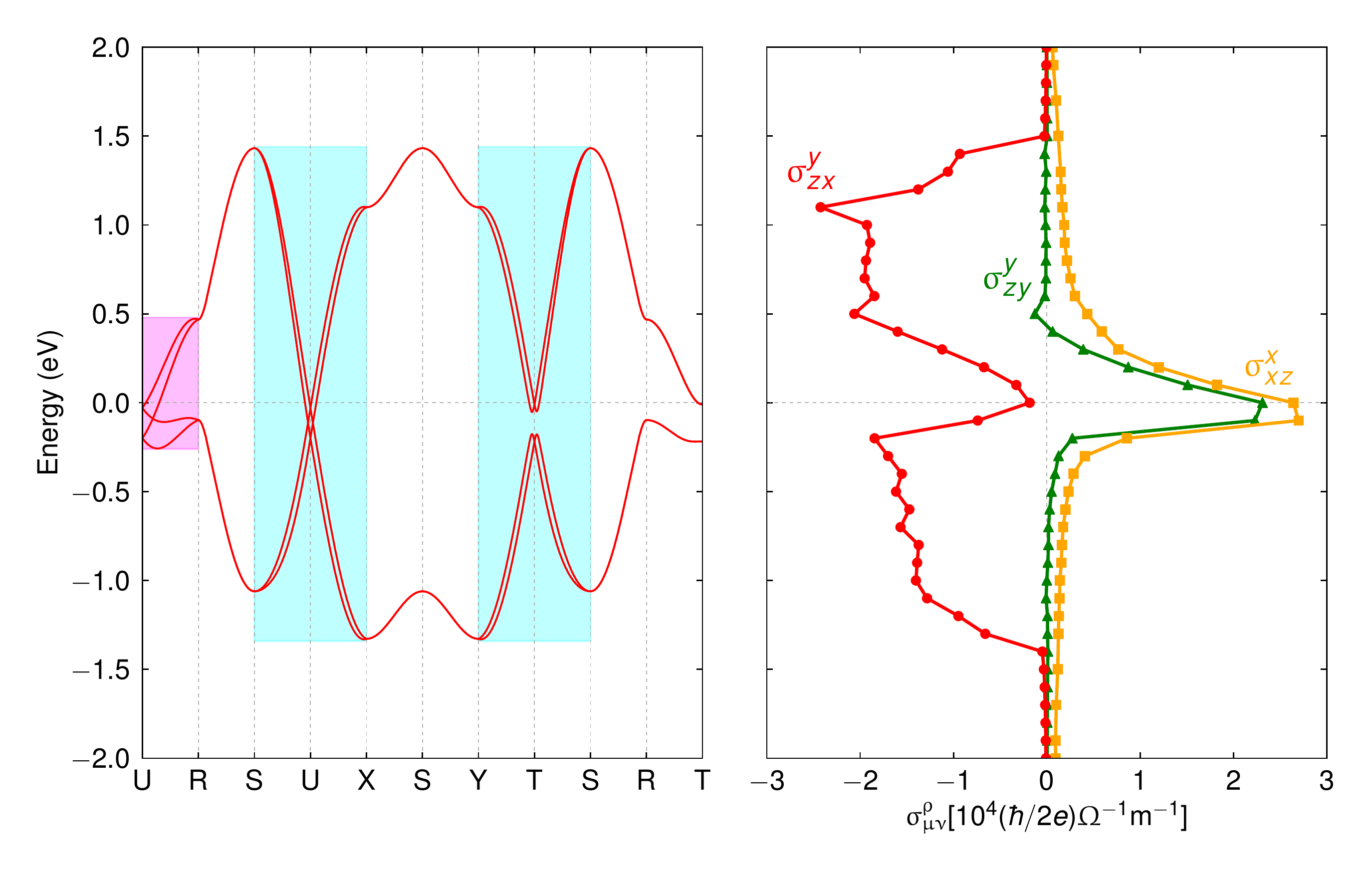}
\caption{Electron energy band structure and spin Hall conductivity \textcolor{black}{in the bulk system.}
The magenta and cyan boxes highlight the nearly degenerate bands that extend from the nodal ring structure.
Spin Hall conductivity $\sigma_{\mu\nu}^{\rho}$ is presented as a function of the Fermi level $\epsilon_F$ (vertical axis) for three configurations in which the system shows the largest response: $\sigma_{zx}^{y}$ (red dot), $\sigma_{zy}^{y}$ (green triangle), $\sigma_{xz}^{x}$ (orange square).
Here, $x$, $y$, $z$ represent the pseudo-cubic axes of the system.
}
\label{fig:band-SHC}
\end{figure*}

Now we consider the commutation relations of the little group elements listed in Table \ref{tab:littlegroup}.
By using Eq. \ref{eq:symm-transformation}, we can find the commutation relations for the U point and RS and SX lines as follows.
\begin{equation}
\begin{array}{ll}
(\textup{U}) & [G_n,G_b]=[G_n,S_a]=\{G_b,S_a\}=0,
\\
(\textup{RS}) & \{G_n,G_b\}=0,
\\
(\textup{SX}) & \{G_n,S_a\}=0.
\end{array}
\label{eq:commutation}
\end{equation}
The anti-commutation relations impose constraints on electron band structure: energy levels at the high symmetry point and lines must be at least fourfold-degenerate due to the anti-commutativeness and the Kramers degeneracy.
The minimal fourfold degeneracy is actually observed in the band structure shown in Fig. \ref{fig:band-SHC} (two fourfold-degenerate bands at the U point and along the RS and XS lines).

More importantly, the anti-commutation relations determine the $G_n$-eigenvalue structure within each fourfold-degenerate energy level.
As illustrated in Fig. \ref{fig:nodalline}, the two levels at the U point are characterized by the different eigenvalues: $n_+$ for the upper and $n_-$ for the lower.
{The upper level consists of four states $\{|n_+,a_+\rangle$, $|n_+,a_-\rangle$, $\Theta |n_+,a_+\rangle$, $\Theta |n_+,a_-\rangle\}$ which are simultaneous eigenstates of $G_n$ and $S_a$.
Here $\Theta$ is the product of time-reversal and spatial-inversion, and it satisfies $[\Theta,G_n]=0$.
These states form a four dimensional representation with the little group structure stated in Eq. \ref{eq:commutation}.
In this representation, both $a_+$ and $a_-$ eigenstates are required by the relation $\{G_b,S_a\}=0$; under the $G_b$ operation, $a_+$ state is mapped into $a_-$ state and vice versa ($|a_+\rangle \overset{G_b}{\leftrightarrow} |a_-\rangle$).
Similarly, the lower level is formed by four states $\{|n_-,a_+\rangle$, $|n_-,a_-\rangle$, $\Theta |n_-,a_+\rangle$, $\Theta |n_-,a_-\rangle\}$ which realize another four dimensional representation of the little group.
However, there is no symmetry requirement that $n_+$ and $n_-$ eigenstates must coexist in each of the two energy levels at the U point.
}

{
Along the SX line, each level comprises four states $\{|n_+\rangle$, $|n_-\rangle$, $\Theta|n_+\rangle$, $\Theta|n_-\rangle\}$ due to the relation $\{G_n,S_a\}=0$ ($\Rightarrow |n_+\rangle \overset{S_a}{\leftrightarrow} |n_-\rangle$).}
Here we stress that both $n_{+}$ and $n_{-}$ eigenstate appear in each level, in contrast to the case at the U point.
This means that there must be $G_n$-partner exchange between the upper and lower levels and thereby band crossing between the two bands involved in the partner exchange, along any path from the U point to the SX line (see Fig. \ref{fig:nodalline}).
Similar argument works for any path connecting the U point with the RS line, in which case the relation $\{G_n,G_b\}=0$ leads to the coexistence of $n_{+}$ and $n_{-}$ states in each of two energy levels.
Consequently, band crossing occurs along a ring centered around the U point, and the nodal ring is protected so long as the nonsymmorphic symmetries $\{G_n$,$G_b$,$S_a\}$ are preserved in the system. 

Although breaking of the $n$-glide symmetry gaps out the nodal ring, the presence of additional nonsymmorphic symmetries can tune the nodal ring into {Weyl or} Dirac point nodes.\cite{Chen_2015, AV_2015} This situation occurs when the $n$-glide and $m$ symmetries are broken while preserving the $b$-glide symmetry. To completely gap out the nodal ring at all points, requires both the $n$-glide and $b$-glide symmetries to be broken. In the next sections, we discuss the spin Hall effect in the bulk and film systems.

\section{Spin Hall effect in bulk system} \label{shc_bulk}

Intrinsic spin Hall effect in SrIrO$_3$ is investigated with a linear response theory.
We compute the spin Hall conductivity (SHC) tensor $\sigma_{\mu\nu}^{\rho}$ using the Kubo formula:\cite{Sinova2015,Guo2008}
\begin{equation}
\sigma_{\mu\nu}^{\rho}
 =  \frac{2e\hbar}{V} \sum_{\bf k} \sum_{\epsilon_{n{\bf k}} < \epsilon_F < \epsilon_{m{\bf k}}}
\textup{Im} \left[ \frac{\langle m {\bf k}| \mathcal{J}_{\mu}^{\rho} | n {\bf k} \rangle \langle n {\bf k}| J_{\nu} | m {\bf k} \rangle }{(\epsilon_{m{\bf k}}-\epsilon_{n{\bf k}})^2} \right] .
\label{eq:SHC}
\end{equation}
In this expression, $J_{\nu} ~(= \sum_{\bf k} \psi_{\bf k}^{\dagger} \frac{\partial H_{\bf k}}{\partial k_{\nu}} \psi_{\bf k})$ is the charge current,
and
$\mathcal{J}_{\mu}^{\rho} ~(= \frac{1}{4} \{ \sigma^{\rho} , {J}_{\mu} \})$ is the spin current with the $j_{\textup{eff}}=1/2$ spin represented by the Pauli matrix $\sigma^{\rho}$.
Other quantities in the expression are: the volume $V$ of the system, Bloch state $|n{\bf k}\rangle$ with energy $\epsilon_{n{\bf k}}$, and Fermi level $\epsilon_F$.
The spin Hall conductivity connects an applied electric field $E^{\nu}$ with an induced transverse spin current by the relationship $\langle \mathcal{J}_{\mu}^{\rho} \rangle = \sigma_{\mu\nu}^{\rho} E^{\nu}$.
Here, the three indices imply the direction of the applied electric field or charge current ($\nu$), the direction of the induced spin current ($\mu$), and the spin polarization axis of the spin current ($\rho$).
The spin Hall conductivity can be recast as $\sigma_{\mu\nu}^{\rho}=\sum_{n,{\bf k}} [\Omega_{\mu\nu}^{\rho}]_{n{\bf k}}~f_{n{\bf k}}$ with the spin-Berry curvature $[\Omega_{\mu\nu}^{\rho}]_{n{\bf k}}$ and the electron occupation number $f_{n{\bf k}}$ at energy level $\epsilon_{n{\bf k}}$.\cite{Sinova2015,Guo2008}
It is the spin-Berry curvature that generates the transverse spin current as an intrinsic effect of electron band structure (analogous to the Berry curvature in anomalous Hall effect).
\textcolor{black}{However, in contrast to the anomalous Hall effect, the spin Hall conductivity is even under time reversal, and so the spin Hall effect does not require time reversal to be broken.}
In our calculations, the configuration $\{\rho,\mu,\nu\}$ representing the spin Hall effect geometry is specified using the pseudo-cubic axes $\{{\bf x}=({\bf a}-{\bf b})/2,~{\bf y}=({\bf a}+{\bf b})/2,~{\bf z}={\bf c}/2\}$ rather than the orthorhombic lattice vectors $\{{\bf a},{\bf b},{\bf c}\}$. By employing the pseudo-cubic axes, it enables easier visualization of the symmetry transformations of the spin Hall conductivity tensor. In Table \ref{tab:pseudocubic}, we present the correspondence between the pseudo-cubic and orthorhombic axes.

\begin{table}[h]
\caption{Correspondence between the pseudo-cubic (c) and orthorhombic axes (o).}
\begin{tabular*}{\hsize}{@{\extracolsep{\fill}}cc}
\hline
\hline
Pseudo-cubic $\{\bf x,y,z\}$ & Orthorhombic $\{\bf a,b,c\}$
\\
\hline
$[100]_c$ & $[1\textup{-}10]_o$
\\
$[010]_c$ & $[110]_o$
\\
$[001]_c$ & $[001]_o$
\\
\hline
\hline
\end{tabular*}
\label{tab:pseudocubic}
\end{table}

We present the spin Hall conductivity $\sigma_{\mu\nu}^{\rho}$ in Fig. \ref{fig:band-SHC} (right) as a function of the Fermi level $\epsilon_F$.
Our results show unexpectedly large spin Hall conductivity of the order of $10^4(\hbar/e)(\Omega\textup{m})^{-1}$ in the three configurations, $\sigma_{zx}^{y}$, $\sigma_{zy}^{y}$, $\sigma_{xz}^{x}$.
To be specific, $\sigma_{zx}^{y}$ (red dot) has large values over an extended region except around the zero Fermi energy, while $\sigma_{zy}^{y}$ (green triangle) and $\sigma_{xz}^{x}$ (orange square) peak around the zero energy. The zero energy is of particular interest as it corresponds to {the electron filling of the bulk system (i.e., half filled $j_{\textup{eff}}=1/2$ electron energy bands).}
By comparing the spin Hall conductivity with the electron band structure on the left of Fig. \ref{fig:band-SHC}, one can notice that large values of $\sigma_{zx}^{y}$ arise in the energy range that the aforementioned nearly degenerate bands (cyan) are extended over.
On the other hand, the peaks of $\sigma_{zy}^{y}$ and $\sigma_{xz}^{x}$ occur within the energy band width at the UR line (magenta).
This implies that the nearly degenerate bands forming the nodal ring structure are closely related to large spin Hall effect in the system.

\begin{figure*}[t]
\centering
\includegraphics[width=0.97\linewidth]{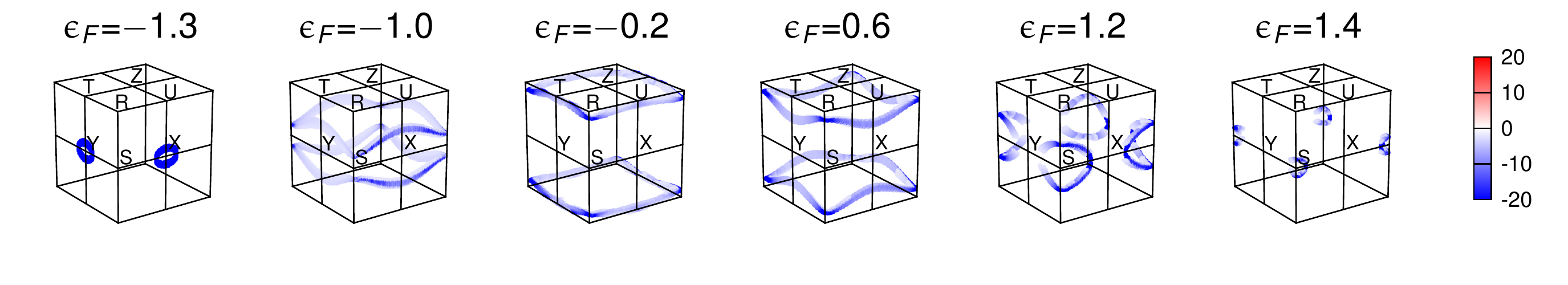}
\caption{Momentum-resolved SHC $\Omega_{zx}^y({\bf k})$ as a function of the Fermi level $\epsilon_F$.}
\label{fig:k-res-SHC-x-zy}
\end{figure*}

\begin{figure}[t]
\centering
\includegraphics[width=0.97\linewidth]{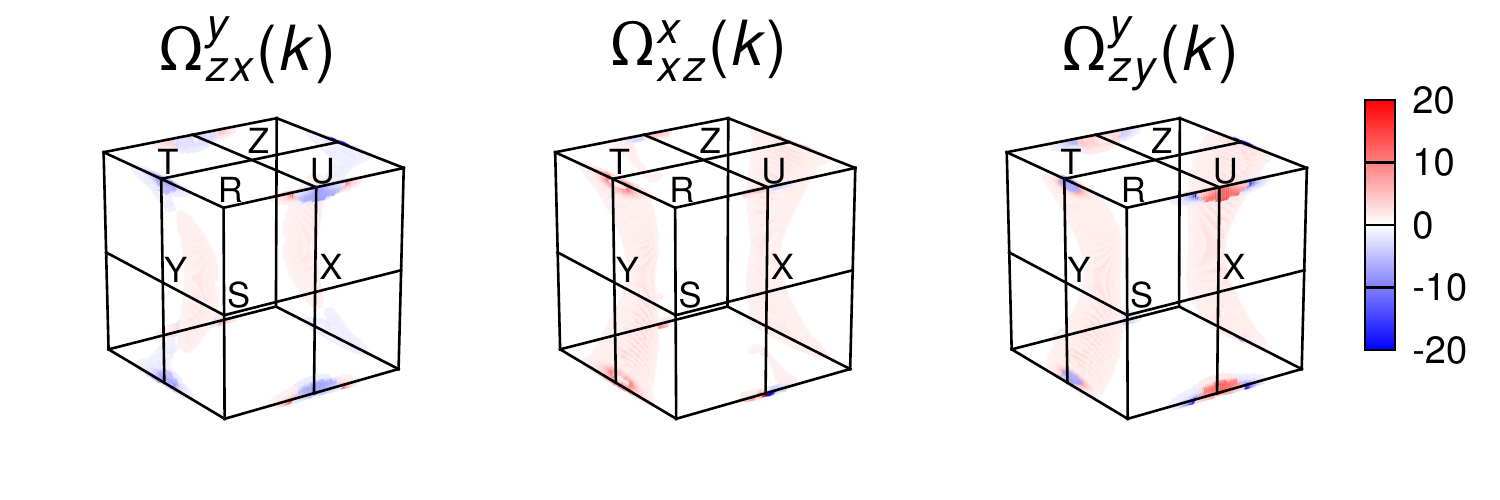}
\caption{Momentum-resolved SHC at $\epsilon_F=0$ for the three configurations in Fig. \ref{fig:band-SHC}.}
\label{fig:k-res-SHC-zero-energy}
\end{figure}

\subsection{Electronic Origin of Large Bulk SHC}
Momentum-resolved spin Hall conductivity provides further useful information about the origin of the large spin Hall effect.
Defining momentum-resolved spin Hall conductivity $\Omega_{\mu\nu}^{\rho}({\bf k})$ by $\sigma_{\mu\nu}^{\rho}=\sum_{\bf k} \Omega_{\mu\nu}^{\rho}({\bf k})$, we investigate the distribution of $\Omega_{\mu\nu}^{\rho}({\bf k})$ in the Brillouin zone.
First, we find that high intensity of $\Omega_{zx}^{y}({\bf k})$ appears in the form of loops extended over the $k_a=\pi$ and $k_b=\pi$ planes as shown in Fig. \ref{fig:k-res-SHC-x-zy}.
The ${\bf k}$-points of the loops have the same sign of $\Omega_{zx}^{y}({\bf k})$, thus leading to constructive contributions to $\sigma_{zx}^{y}$. Remarkably, these loops correspond to the $\bf{k}$-points where the Fermi level crosses the nearly degenerate bands in the $k_a=\pi$ and $k_b=\pi$ planes [compare the resemblance of the Fermi \textcolor{black}{surface cross sections within} the $k_b=\pi$ plane in Fig. \ref{fig:structure} to $\Omega_{zx}^{y}({\bf k})$ in Fig. \ref{fig:k-res-SHC-x-zy}]. This suggests that the `active' electronic states that are contributing the most to $\Omega_{zx}^{y}({\bf k})$ are in fact the states residing in the nearly degenerate band structure in the $k_a=\pi$ and $k_b=\pi$ planes. Therefore, one can conclude that the large values of $\sigma_{zx}^{y}$ indeed originates from the interband transition between the aforementioned nearly degenerate bands in the $k_a=\pi$ and $k_b=\pi$ planes (shaded in Fig. \ref{fig:band-SHC}).

The different behaviors of $\sigma_{zx}^{y}$, $\sigma_{zy}^{y}$, $\sigma_{xz}^{x}$ around the zero Fermi energy (Fig. \ref{fig:band-SHC}) can also be understood by investigating the corresponding $\Omega_{\mu\nu}^{\rho}({\bf k})$. As shown in Fig. \ref{fig:k-res-SHC-zero-energy}, spin Hall conductivity at $\epsilon_F=0$ eV has major contributions broadly from two different parts of the Brillouin zone: (i) a region around the U point and (ii) arch-shaped interior regions connected to the T point. 
Specifically, in the case of $\Omega_{zx}^{y}({\bf k})$, the two parts have a different sign and their contributions to spin Hall conductivity almost cancel each other, which results in highly suppressed $\sigma_{zx}^{y}$ at $\epsilon_F=0$ eV.
However, the other two cases $\Omega_{zy}^{y}({\bf k})$ and \textcolor{black}{$\Omega_{xz}^{x}({\bf k})$} have an almost uniform sign over the two parts, thereby $\sigma_{zy}^{y}$ and $\sigma_{xz}^{x}$ have a large value at $\epsilon_F=0$ eV.

\subsection{Signature of Nonsymmorphic Symmetries in Bulk SHC}
Nonsymmorphic symmetries of the bulk system provide useful constraints on the spin Hall conductivity tensor.
Specifically, one can easily check that \textcolor{black}{$\sigma_{zx}^y=-\sigma_{zy}^x$, $\sigma_{zy}^y = - \sigma_{zx}^x$, and $\sigma_{xz}^x=-\sigma_{yz}^y$} (where $x$ and $y$ are interchanged with an additional minus sign) by each of the $n$-glide, $b$-glide, $a$-screw, and $b$-screw symmetries of the {\it Pbnm} space group.
Hence, the three configurations shown in Fig. \ref{fig:band-SHC} are symmetry-related to other three.
Except the six configurations, we find subdominant spin Hall conductivity, at least one order of magnitude smaller.

Our results show that the system exhibits large spin Hall response when the spin current is induced along the $z$ axis ($\sigma_{zx}^y$ and $\sigma_{zy}^y$ in Fig. \ref{fig:band-SHC}). 
Keeping the spin current direction along the $z$ axis, we change the applied electric field direction to investigate the field direction dependence of spin Hall conductivity.
Specifically, we consider the following configuration.
\begin{equation}
\begin{array}{lcc}
\nu & \parallel & \hat{x}~\textup{cos}\theta + \hat{y}~\textup{sin}\theta,
\\
\mu&\parallel& \hat{z},
\\
\rho&\parallel& \hat{x}~\textup{sin}\theta - \hat{y}~\textup{cos}\theta.
\end{array}
\end{equation}
Here, the field direction ($\nu$) is changed within the $xy$ plane by the angle $\theta$ from the $x$ axis with keeping the three directions $\{\rho,\mu,\nu\}$ orthogonal.
In this setting, by the nonsymmorphic symmetries, the spin Hall conductivity is simply a combination of \textcolor{black}{$\sigma_{zx}^y$} and $\sigma_{zy}^y$:
\textcolor{black}{
\begin{equation}
\sigma_{\mu\nu}^{\rho}=-\sigma_{zx}^y-\sigma_{zy}^y~\textup{sin}2\theta .
\label{eq:field-direction-dep-SHC}
\end{equation}
}
\begin{figure}[t]
\centering
\includegraphics[width=0.94\linewidth]{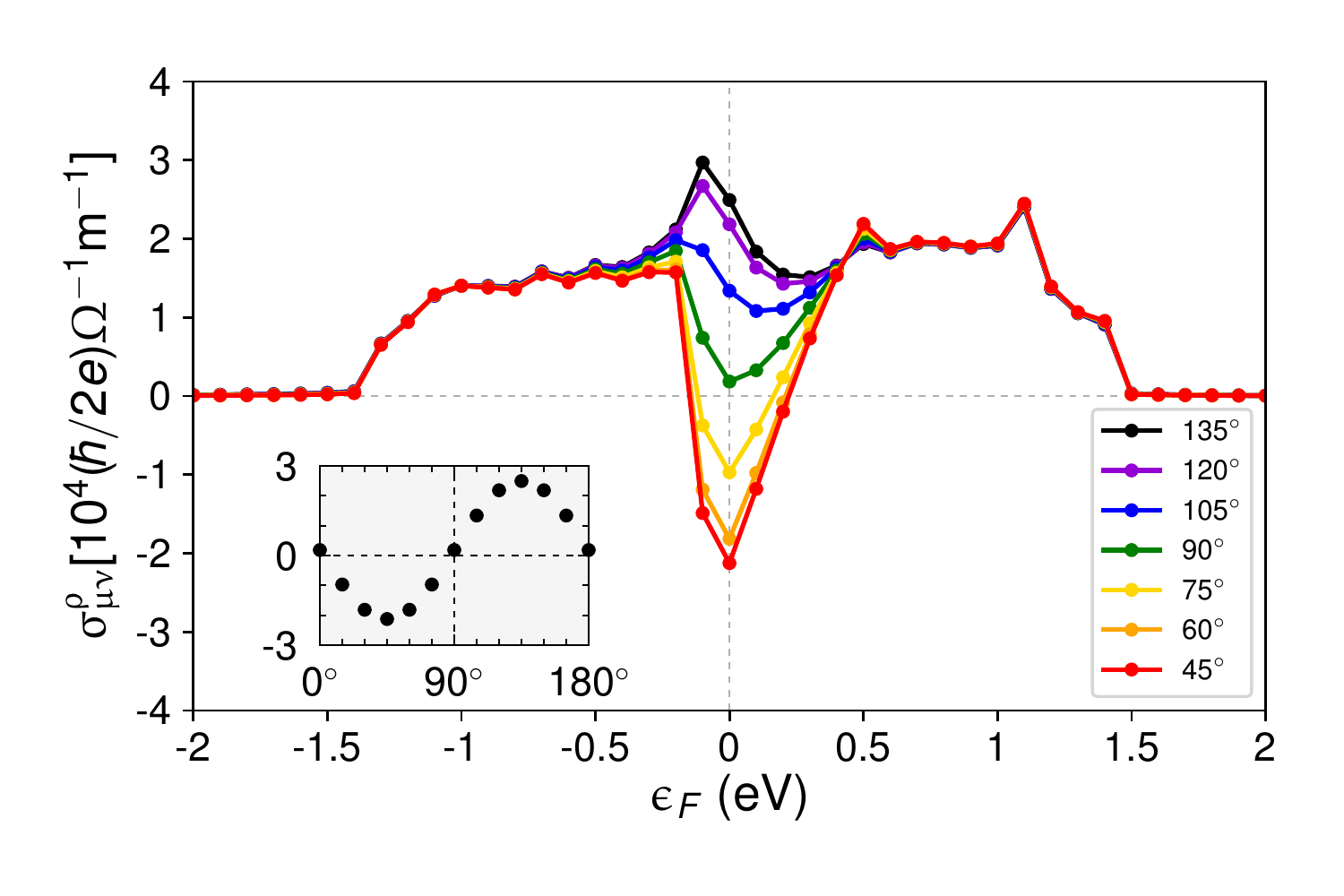}
\caption{Field direction dependence of spin Hall conductivity. Inset: angular dependence of the spin Hall conductivity at the zero Fermi energy.}
\label{fig:direction-dep-SHC}
\end{figure}

As described in Fig. \ref{fig:direction-dep-SHC}, the spin Hall conductivity drastically changes around the zero Fermi energy as the field direction varies.
The largest magnitude of spin Hall conductivity occurs when $\theta=45^{\circ}$ and $\theta=135^{\circ}$ which correspond to the $[110]_c$ and $[1\bar{1}0]_c$ pseudo-cubic axes for the field direction, respectively (those two axes are identical to the $[100]_o$ and $[0\bar{1}0]_o$ orthorhombic axes).
For the experiments on the bulk system, this suggests that the largest spin Hall response is expected for the electric field along the $[100]_o$ or $[0\bar{1}0]_o$ axis.
On the other hand, the characteristic sinusoidal angular dependence (inset of Fig. \ref{fig:direction-dep-SHC}) can be used to experimentally probe the presence of the nonsymmorphic symmetries in the system.
The sinusoidal behavior described in Eq. \ref{eq:field-direction-dep-SHC} is dictated by the {\it Pbnm} nonsymmorphic symmetries regardless of the details of the Hamiltonian.

\begin{figure}[b]
\centering
\includegraphics[width=1.0\linewidth]{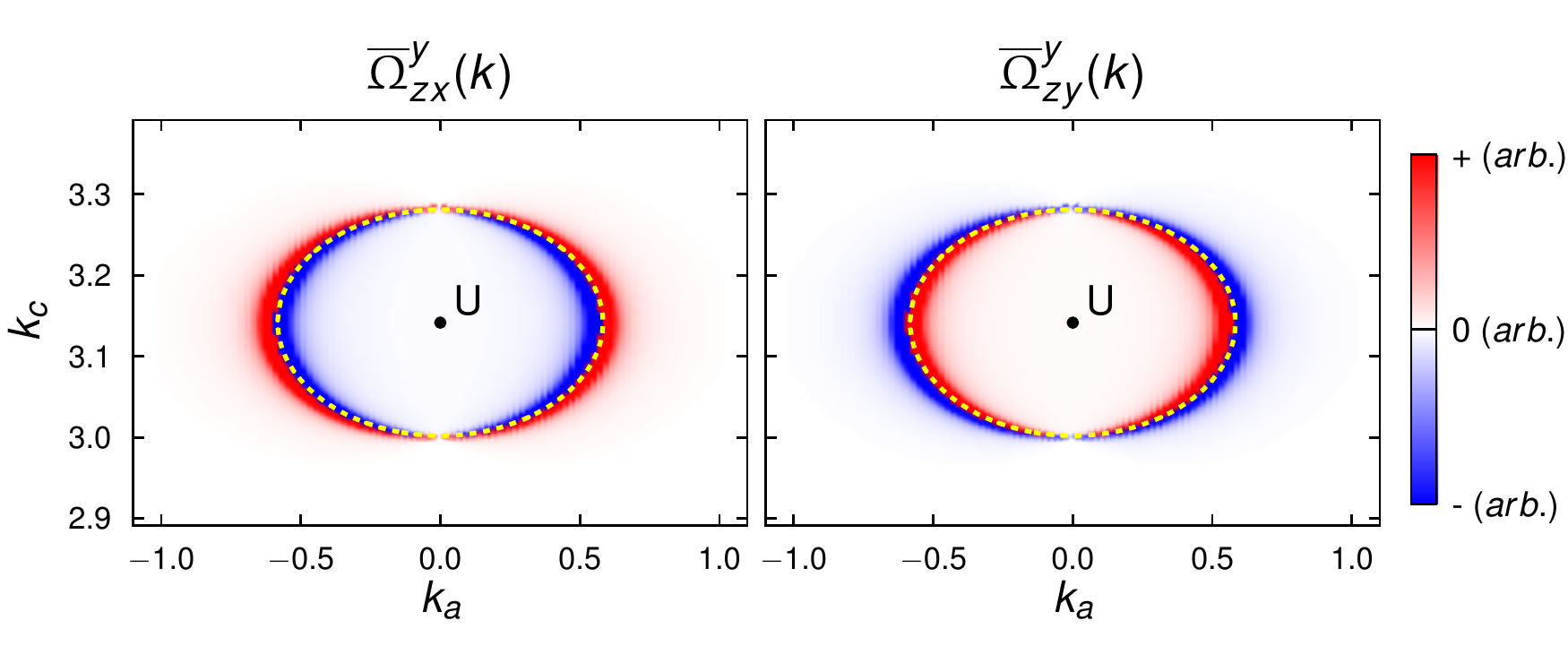}
\caption{
{Spin-Berry curvature of the nodal ring band structure.
The color maps depict the net spin-Berry curvature ${\overline{\Omega}}_{\mu\nu}^{\rho}({\bf k})(=\sum_{n=1}^4 [\Omega_{\mu\nu}^{\rho}]_{nk}$) calculated for the lowest two doubly degenerate bands in Figs. \ref{fig:structure} (c) and \ref{fig:band-SHC} on the $k_b=\pi$ plane.
The yellow dashed line denotes the nodal ring {residing in the $k_b = \pi$ plane.}}
}
\label{fig:nodal-sign}
\end{figure}

\subsection{Role of Nodal Ring in Bulk SHC}

{
Now we examine the spin-Berry curvature contained in the nodal ring band structure and its contribution {to} the bulk SHC.
Fig. \ref{fig:nodal-sign} {depicts} the spin-Berry curvature of the nodal ring band structure.
High intensity spin-Berry curvature is found near the nodal ring band crossing.
Moreover, the sign of the spin-Berry curvature changes across the nodal ring in the $k_b=\pi$ plane {(}such sign-changing behavior of high intensity spin-Berry curvature across band crossing points {has also} been found in recent {\it ab initio} studies on Weyl semimetals \cite{TaAs_SHE_2016}{)}.
One interesting feature here is that the spin-Berry curvature is nonzero only at the $k_b=\pi$ plane, and it immediately vanishes off the plane.}

{
Although the nodal ring structure generates high intensity of spin-Berry curvature {(near the ring)}, its net contribution to the {total bulk} SHC {is quite small} due to the massive cancellation between the opposite signs of the spin-Berry {curvature} in the $k_b=\pi$ plane.
{We directly examine this large cancellation by focusing on the nodal ring's isolated contributions (i.e., from the $\bf{k}$-points in the rectangular region just-enclosing the nodal ring in the $k_b=\pi$ plane). For the three configurations, $\sigma_{zx}^{y}$, $\sigma_{zy}^{y}$ and $\sigma_{xz}^{x}$, the isolated contribution (weighted by the entire Brillouin zone contribution) of the nodal ring is $6\%$, $0.7\%$, and $0.03\%$ respectively for $\epsilon_F=0$ eV, and $7\%$, $2\%$, and $0.01\%$ respectively for $\epsilon_F=-0.1$ eV. Furthermore, we also investigated the impact of the size of the nodal ring (controlled by the spin-dependent hopping parameter $t_d^o$) on the SHC. Our computations determined that despite the size of the nodal ring increasing by two-fold, four-fold or even six-fold, the change in the SHC is very small and not substantial (order of magnitude remains the same).
This suggests that any contributions arising from and about the the nodal ring are promptly cancelled, resulting in its benign contribution to the SHC.}}

\section{Impact of Breaking Nonsymmorphic Symmetries on Bulk SHC} \label{shc_bulk_break}

In a realistic system, it is highly probable (and likely) that one or many of nonsymmorphic symmetries can be broken. The breaking of these symmetries (such as the glide symmetries) can be the result of growth defects, impurities, or external strain/pressure on the bulk. This can lead to, depending on which symmetry is broken, the lifting of degeneracy at certain high symmetry locations in the Brillouin zone or (even more drastically) the removal of certain topological features in the band structure like the nodal ring. In fact, the gapping of the nodal ring is not a rare occurrence in thin film configurations of SrIrO$_3$ as has been observed through ARPES studies on epitaxial thin films,\cite{Shen_gap} as well as in epitaxially strained thin films.\cite{AV_2015} One way to incorporate such a broken symmetry is a perturbation term in the bulk model:
\begin{equation}
h_{gap} = t_{gap} \Big[ \sin(k_y) - \sin(k_x) \Big] \sin(k_z) \nu_y \tau_y.
\end{equation}

\begin{figure}[t]
\centering
\includegraphics[width=0.84\linewidth,angle=0]{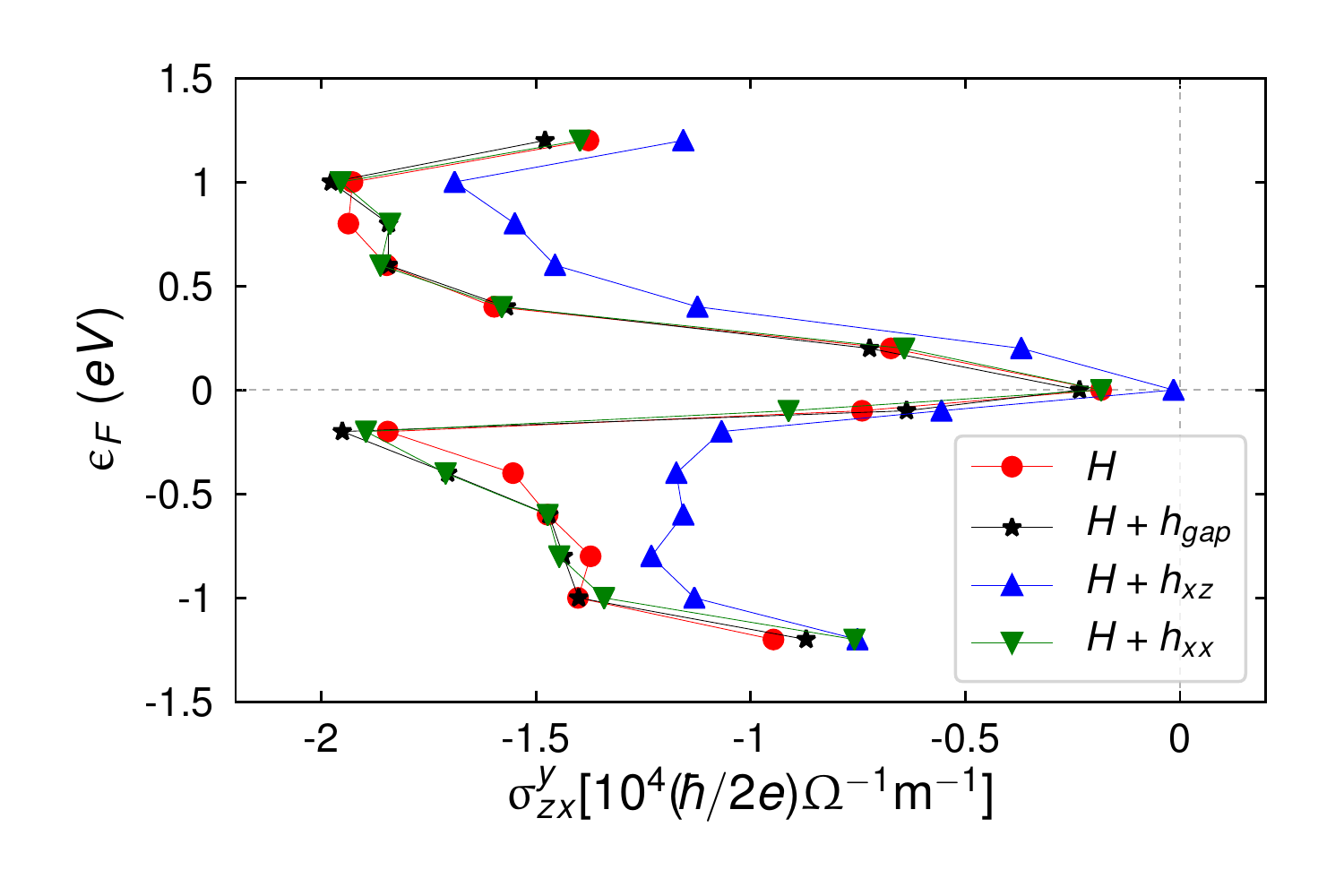}
\includegraphics[width=0.84\linewidth,angle=0]{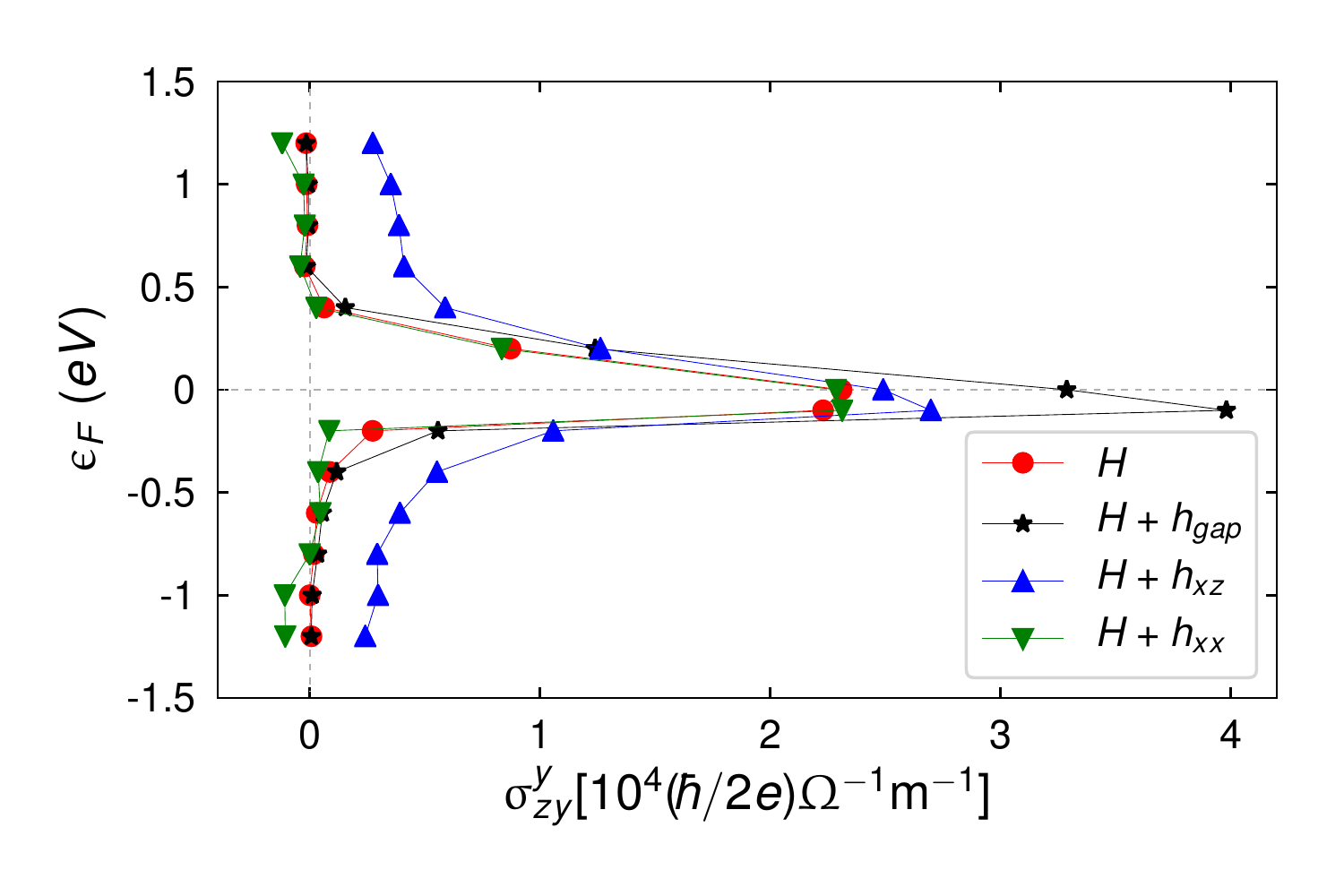}
\includegraphics[width=0.84\linewidth,angle=0]{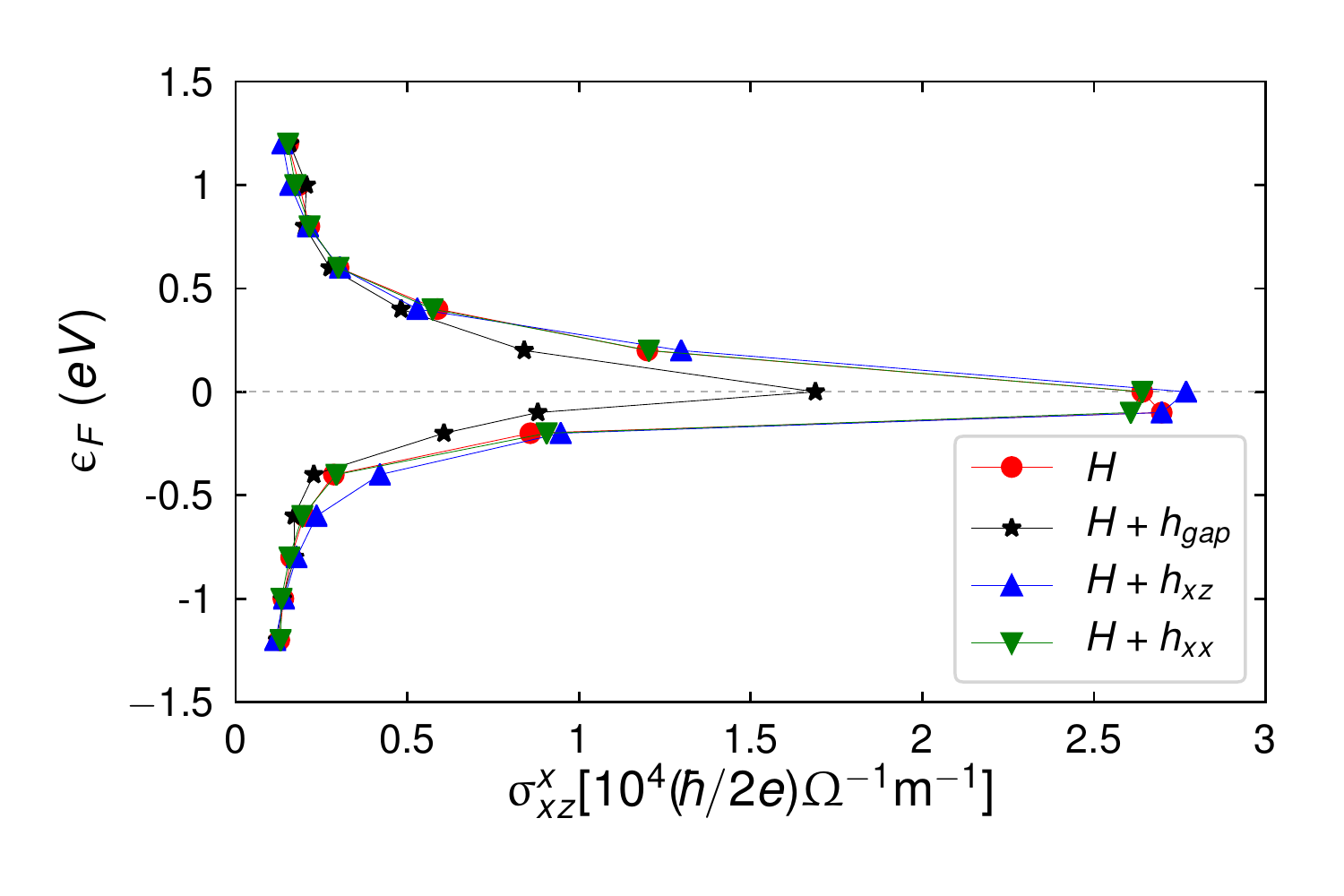}
\caption{SHC of original bulk model (${H}$) compared to bulk model augmented by various perturbations.}
\label{fig:SHC_perturb}
\end{figure}

This term preserves the inversion, time-reversal, and mirror symmetries. However, it breaks the $b$-glide and $n$-glide symmetries, and so it meets the required criterion to completely gap out the nodal ring. {This can be seen by the commutation of $h_{gap}$ and the symmetry operators (Eq. \ref{eq:symm-transformation} and Eq. \ref{eq:symm-transformation-2} in Appendix \ref{Perturb_op}).} Moreover, this term involves the in-plane and sublayer degrees of freedom ($\tau$ and $\nu$, respectively) and so physically this mimics a possible physical strain on the system that has led to the breaking of these symmetries. The choice of $t_{gap}=0.01eV$ is at least an order of magnitude smaller than the other energy scales in the system, and thus it acts as a small perturbation. Furthermore, this term introduces a Dirac mass gap of $\approx 40$meV, consistent with experimental observations of gapped {Dirac points in SrIrO$_3$ thin films}.\cite{Shen_gap} Fig. \ref{fig:gapped_together} in Appendix \ref{app_perturbations} depicts the impact of the perturbative term $h_{gap}$ on the band structure (along high symmetry directions {U$\rightarrow$R and U$\rightarrow$X}).

However, $h_{gap}$ is only one of a myriad of possible symmetry breaking terms that could be introduced, involving again only the in-plane and sublayer degrees of freedom. We limit our possibilities to terms that can modify the band structure in regions of the Brillouin zone that contribute the most to the spin Hall conductivity (highlighted in Fig. \ref{fig:band-SHC}); these are the terms that have the greatest chance in modifying the spin Hall conductivities. In particular, we determined two types of symmetry breaking terms that introduced qualitative changes in the band structure, and thus potentially had the chance to introduce a non-zero change in the spin Hall conductivities:
\begin{equation}
\begin{aligned}
& h_{xx} = t_{xx} \Big[\sin(k_y) - \sin(k_x) \Big] \sin(k_z) \nu_x \tau_x, \\
& h_{xz} = t_{xz} \Big[\sin(k_y) + \sin(k_x) \Big] \sin(k_z) \nu_x \tau_z.
\end{aligned}
\end{equation}

These terms break $n$-glide and $m$ symmetries, while preserving $b$-glide symmetry. The choice of $t_{xx,xz}$=$0.05eV$ was taken as it ensured that these terms act as perturbations on the original bulk model, while also creating a qualitative change in the band structure (Figs. \ref{fig:xx_together}, \ref{fig:xz_together} in Appendix \ref{app_perturbations}). Augmenting the original Hamiltonian with these perturbative terms, we present the spin Hall conductivities of the three configurations for which the system had the greatest response in Fig. \ref{fig:SHC_perturb}. 

\begin{figure}[b]
\centering
\includegraphics[width=0.78\linewidth]{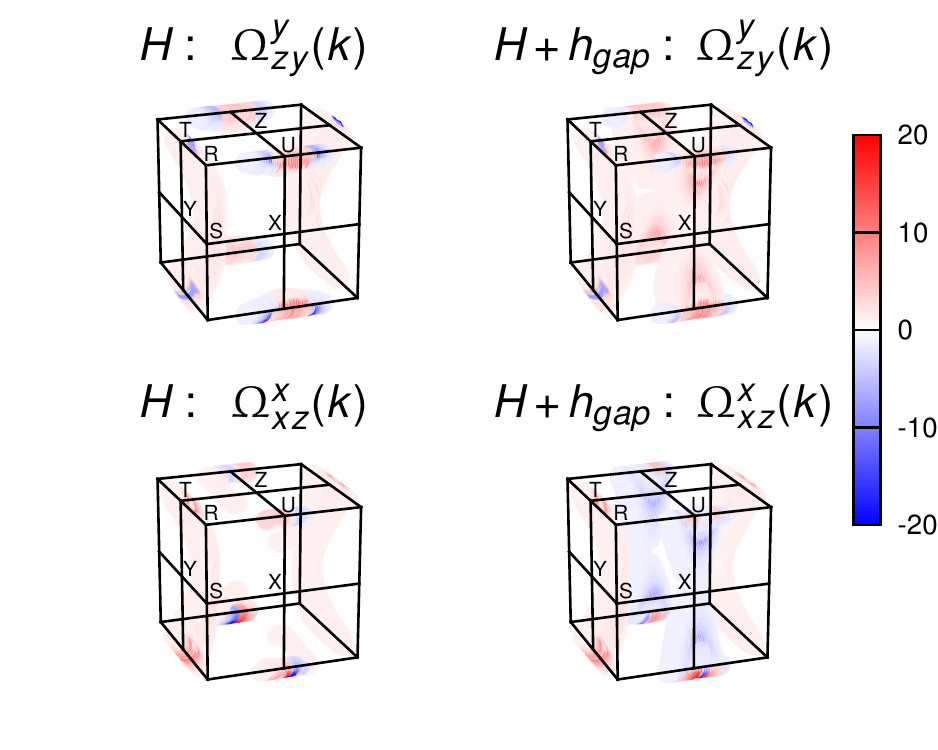}
\caption{Momentum-resolved SHC at the zero Fermi level for the $\sigma_{zy}^y$ and $\sigma_{xz}^x$ depicting the impact of gapping the nodal ring in Fig. \ref{fig:SHC_perturb}.}
\label{fig:berry_gapped_comp}
\end{figure}

\begin{figure*}[t]
\centering
\includegraphics[width=\linewidth]{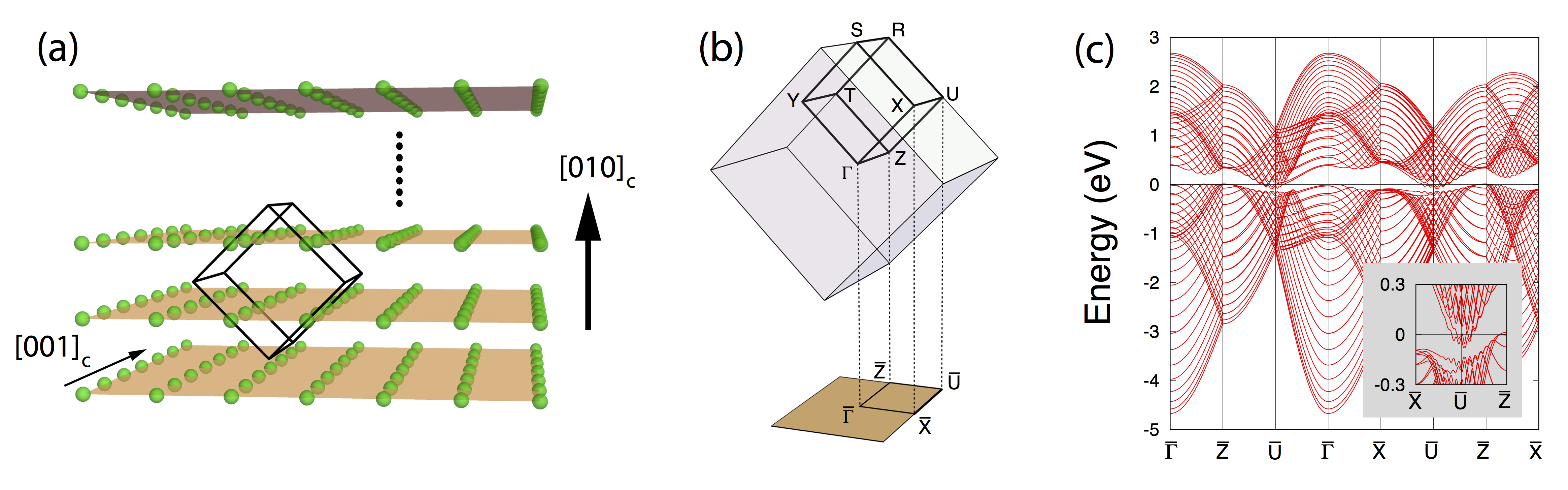}
\caption{Geometry and electronic structure of (010)$_c$ thin film. (a) Lattice geometry of Ir ions (green). The orange planes describe a stacking of (010)$_c$ layers of Ir ions in thin film system. The black lines and arrows represent bulk unit cell and some of the crystallographic axes, respectively. (b) Two dimensional Brillouin zone of (010)$_c$ film. The bulk Brillouin zone (cube) is projected and folded into the film Brillouin zone (bottom rectangle). (c) Electronic energy band structure of (010)$_c$ film with 16 layers of Ir ions. \textcolor{black}{The inset shows the band structure magnified around the $\bar{\textup{U}}$ point.}}
\label{fig:thinfilm-band}
\end{figure*}

Interestingly, despite the inclusion of various symmetry breaking perturbative terms, the component $\sigma_{zx}^y$ remains stable and still maintains the large spin Hall conductivity at all the previously noted Fermi levels. This is not altogether surprising as the greatest contributions to this component arise from the nearly degenerate electron band structure (cyan in Fig. \ref{fig:band-SHC}). Since the perturbative terms $h_{gap}$ and $h_{xx}$ {do not affect} the nearly degenerate regions, the spin Hall conductivity magnitude is preserved with the inclusion of these terms. However, there is a small modification, but remains within the same order of magnitude, in the spin Hall conductivity due to $h_{xz}$. This term marginally changes the band structure in the nearly degenerate regions by only lifting the degeneracy at the band crossing T point. On the other hand, the components $\sigma_{zy}^y$ and $\sigma_{xz}^x$ do see a jump in the spin Hall conductivity by the gapping of the nodal ring close to the zero energy level; apart from $\sigma_{zy}^y$ experiencing a similar shift as $\sigma_{zx}^y$ due to $h_{xz}$, these configurations are stable to the other perturbations. By considering the momentum-resolved SHC at the zero energy level (Fig. \ref{fig:berry_gapped_comp}) one notices new arch-like features that develop about the U point. For $\sigma_{zy}^y$, the new features have the same sign as the arch-like features from the non-perturbed model and so the spin Hall conductivity is enlarged; while for $\sigma_{xz}^x$, the new features have the opposite sign and so diminish the spin Hall conductivity. Nevertheless, the enlargement/diminishment is not substantial and the spin Hall conductivity remains within the same order of magnitude. Hence, the bulk spin Hall conductivity is stable and robust to the introduction of various symmetry breaking terms (other symmetry breaking terms further support this conclusion).

\section{Spin Hall effect in thin film system} \label{shc_film}

Now we turn our attention to thin film systems of SrIrO$_3$.
Recent experiments \cite{SIO_SHE_exp} \textcolor{black}{discovered} that SrIrO$_3$ thin films exhibit surprisingly large spin Hall conductivity [$\sigma_{\textup{SH}}^{\textup{exp}}\sim10^5(\hbar/2e)\Omega^{-1}\textup{m}^{-1}$], which is about one order of magnitude larger than values predicted for the bulk system.
The experimental film samples were grown along the [010]$_c$ direction up to a thickness of 20 bulk unit cells.
Due to the nature of the experimental setup, only the spin current induced along the [010]$_c$ film direction was measured (see Fig. \ref{fig:thinfilm-band}).
This large thin film spin Hall response is unexpected, since our theory for the bulk system predicts rather small responses for the same configuration corresponding to the experiment.

\begin{figure}[t]
\centering
\includegraphics[width=0.99\linewidth]{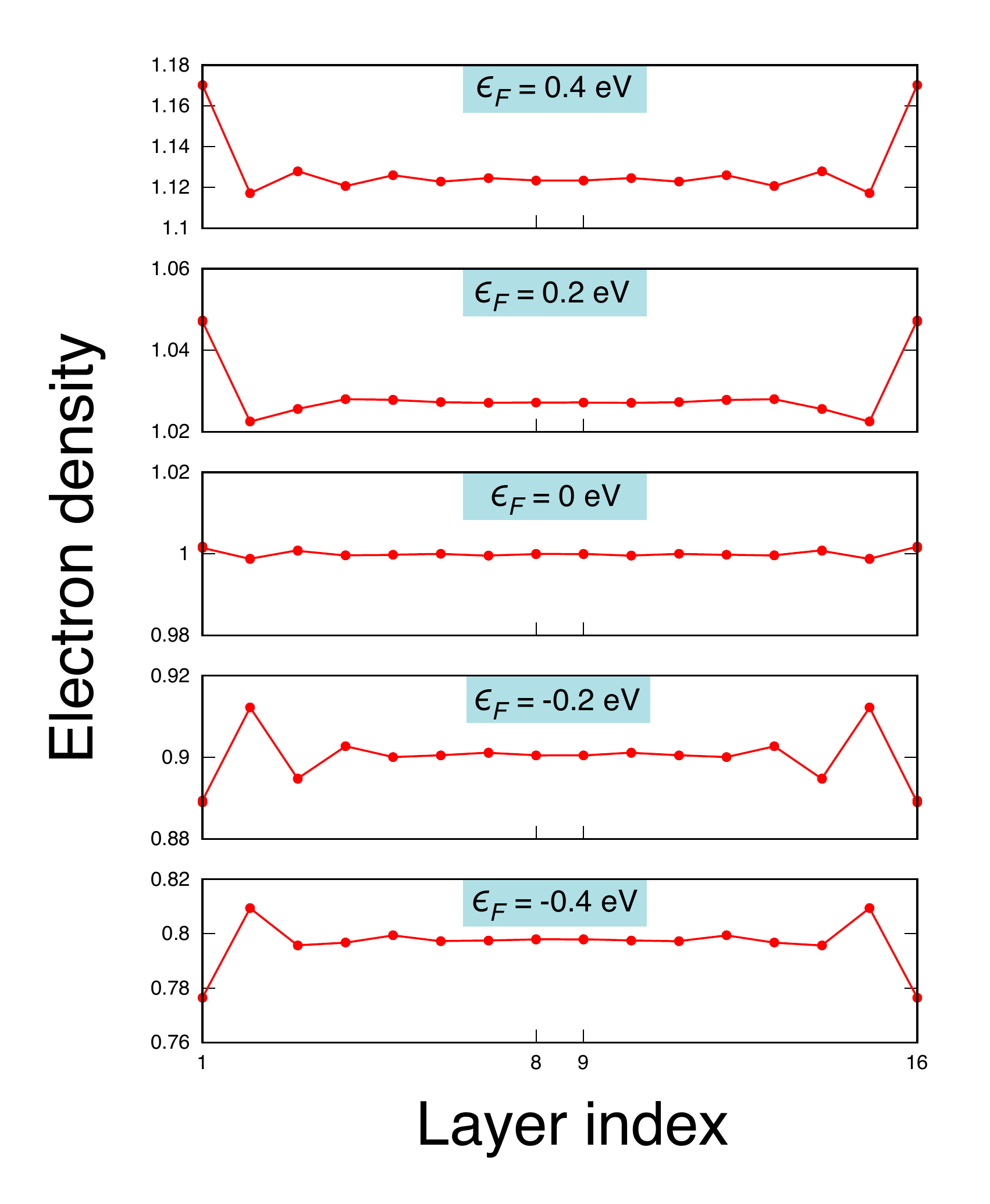}
\caption{
Layer dependence of electron density in the thin film system. The layer index represents the 16 Ir-layers of the film system ($1\sim16$: top to bottom layer). The symmetric pattern of the electron density about the central layers (8,9) is due to the inversion symmetry remaining in the film system.
}
\label{fig:thinfilm-el-density}
\end{figure}

In this section, we study the (010)$_c$ thin film system to get an insight into the large enhancement of SHC in the film geometry.
The (010)$_c$ thin film in our study is described using the bulk tight-binding model $H$ on the slab lattice geometry shown in Fig. \ref{fig:thinfilm-band} (a).
\textcolor{black}{Our calculation results presented below were obtained for the system with 16 layers of Ir sites (8 bulk unit cells along the [010]$_c$ direction) which is the largest thickness that \textcolor{black}{allowed us to} reach convergence in the SHC calculations within a reasonable amount of time}.
The finite thickness of the film system leads to the breaking of some of the original bulk symmetries, for instance the $n$-glide and $b$-glide symmetries as well as the translational symmetry along the [010]$_c$ direction (see Table \ref{tab:Pbmn} for the remaining symmetries in the film system).
As a consequence of the lower lattice symmetry, the nodal line band crossing of the bulk system is gapped out in the (010)$_c$ film [see Fig. \ref{fig:thinfilm-band} (c)].
\textcolor{black}{Moreover, the film system has a nonuniform electron distribution over the layers of the Ir sites as shown in Fig. \ref{fig:thinfilm-el-density} (by contrast, the bulk system has a uniform electron distribution over the four sublattices by lattice symmetry).}

\begin{figure}[t]
\centering
\includegraphics[width=\linewidth]{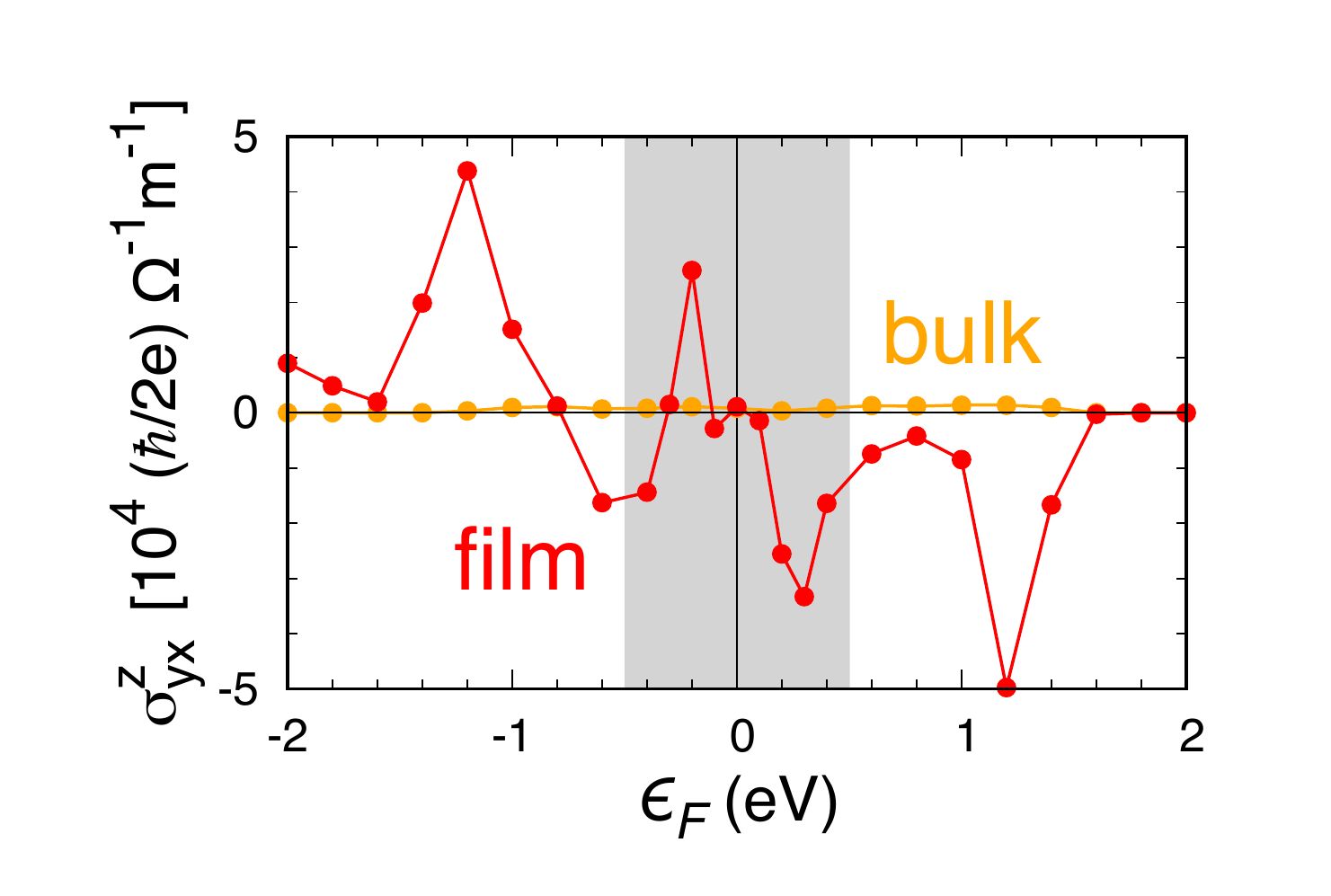}
\caption{Spin Hall conductivity $\sigma_{yx}^z$ of the thin film as a function of Fermi energy $\epsilon_F$. Corresponding bulk SHC is shown for comparison.}
\label{fig:thinfilm-SHC}
\end{figure}

\textcolor{black}{Figure} \ref{fig:thinfilm-SHC} displays the spin Hall conductivity obtained from the linear response theory for the thin film.
Our result shows large film SHC in the configuration corresponding to the experiments: $\sigma_{yx}^z \sim 10^4(\hbar/2e)\Omega^{-1}\textup{m}^{-1}$ around the zero Fermi energy \textcolor{black}{(see the shaded region in \textcolor{black}{Fig. \ref{fig:thinfilm-SHC}})}.
However, in the same configuration, the bulk system shows rather small SHC as shown in the figure (at least one order of magnitude smaller compared to the film SHC).
Thus, natural questions to ask are (i) why is the film's spin Hall response so different from the bulk system and (ii) what is the origin of the large spin Hall response in the film.

\begin{figure*}
\centering
\includegraphics[width=\linewidth]{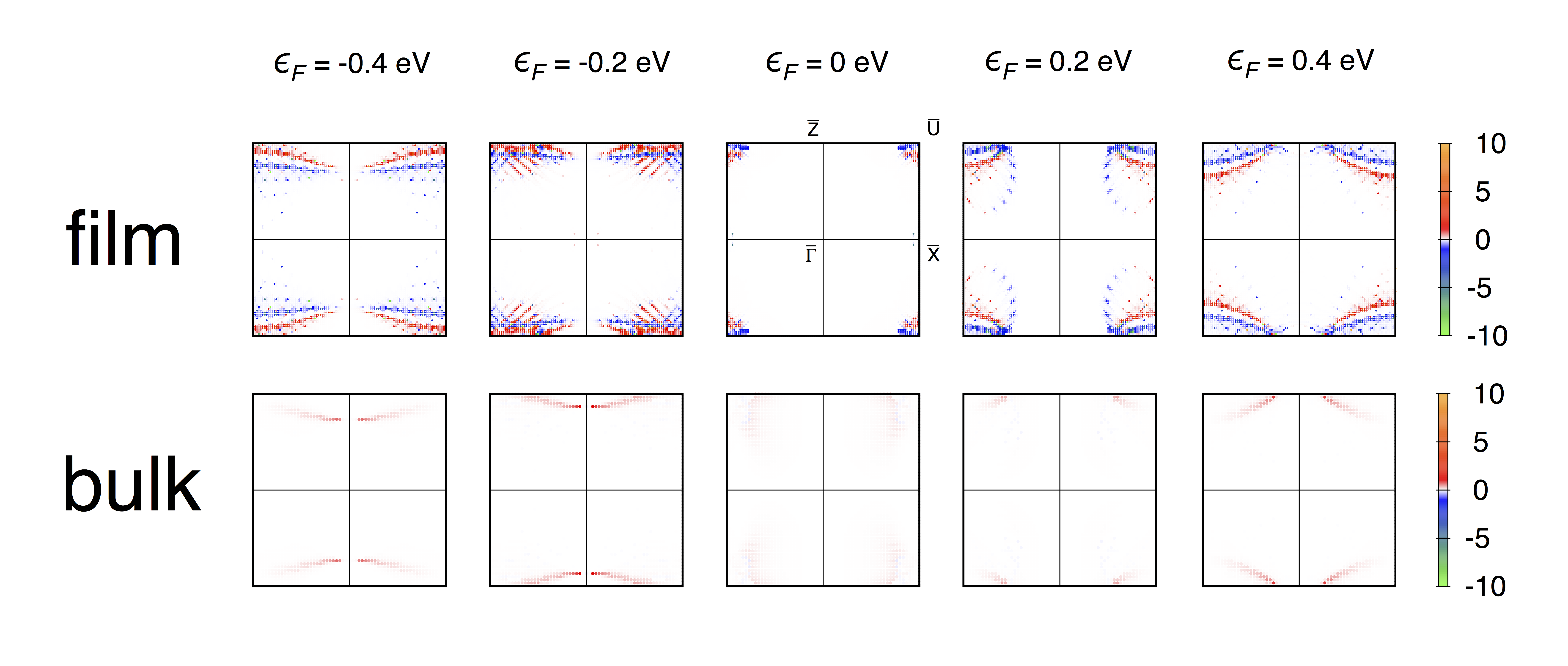}
\caption{Momentum-resolved spin Hall conductivity $\Omega_{yx}^z({\bf k})$ of the thin film (upper) and bulk (lower) systems. The bulk result was obtained by taking into account the zone folding described in Fig. \ref{fig:thinfilm-band} (b) and using the same unit for the film and bulk results for comparison with the film result.}
\label{fig:thinfilm-momresSHC}
\end{figure*}

To understand the difference between the film and bulk and the origin of the large SHC in the film, we resolve the SHC in the two dimensional Brillouin zone of the film as shown in Fig. \ref{fig:thinfilm-momresSHC}.
The upper and lower panels represent the momentum-resolved SHC $\Omega_{yx}^z({\bf k})$ for the film and bulk systems, respectively; the bulk result was obtained by taking into account the zone folding described in Fig. \ref{fig:thinfilm-band} (b).
The similarities and differences between the film and bulk systems are revealed in the distribution of $\Omega_{yx}^z({\bf k})$.
Firstly at each Fermi energy, overall patterns of $\Omega_{yx}^z({\bf k})$ in the film and bulk are quite similar to each other.
Despite this similarity, significant difference in the magnitude and sign of $\Omega_{yx}^z({\bf k})$ is observed between the film and bulk, suggesting that electron wave functions in the thin film significantly deviate from wave functions in the corresponding bulk system. 

\begin{figure}
\centering
\includegraphics[width=0.5\linewidth]{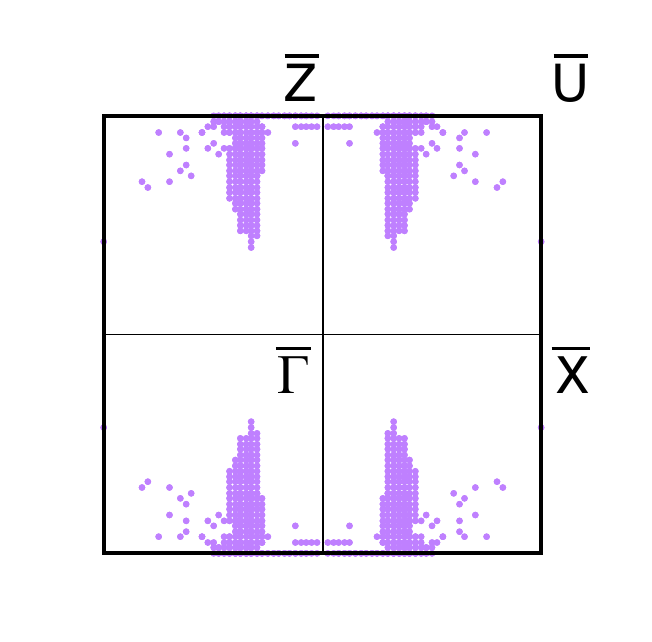}
\caption{
Locations of the surface states in the Brillouin zone.
The purple dots represent the states with electron density with more than 66 percent at the top four and bottom four layers among the entire 16 Ir-layers of the film system.
The surface state locations, which are qualitatively different from the pattern of $\Omega_{yx}^z({\bf k})$ in Fig. \ref{fig:thinfilm-momresSHC}, indicate that the large film SHE is mainly an effect by bulk-like states rather than the surface states in the thin film system.
}
\label{fig:thinfilm-surface-states}
\end{figure}

In addition to the modification of electron wave functions, the surface states arising in the film (which are localized around the boundary surfaces) could be another source for the difference between the film and bulk spin Hall responses.
Fig. \ref{fig:thinfilm-surface-states} depicts the locations at which surface states occur in the Brillouin zone.
Comparing Figs. \ref{fig:thinfilm-momresSHC} and \ref{fig:thinfilm-surface-states}, one observes that the distribution of the surface states in momentum space does not seem \textcolor{black}{to} match well with the pattern of $\Omega_{yx}^z({\bf k})$ in the film system.
This supports the idea that the large film SHC is mainly an effect of bulk-like electron states rather than the surface states. The bulk-like electron states are, however, substantially modified from states in the bulk system
{
as indicated by the electron density (Fig. \ref{fig:thinfilm-el-density}) and spin-Berry curvature distribution (Fig. \ref{fig:thinfilm-momresSHC}).
The change in the wave functions (and subsequent enhancement of the spin Hall effect in the film geometry) is attributed to the lower lattice symmetry in the thin film system. 
In the absence of the film direction translation, $n$-glide, $b$-glide, and other lattice symmetries, electron wave functions are obviously less constrained {as} compared to in the original bulk environment.
}

\section{Conclusions} \label{conclusion}

In this work, we examined the intrinsic spin Hall effect in both bulk and thin film configurations of SrIrO$_3$ using linear response theory. We employed the $j_{\textup{eff}}=1/2$ tight binding model constructed from \textit{ab initio} studies of SrIrO$_3$.\cite{Carter_2012,Chen_2015}
From our bulk SrIrO$_3$ studies in Sec. \ref{shc_bulk} and \ref{shc_bulk_break}, we found an unexpectedly large spin Hall conductivity [$\sigma_{\textup{SH}}\sim10^4(\hbar/e)(\Omega\textup{m})^{-1}$] in three configurations: $\sigma_{zx}^{y}$, $\sigma_{zy}^y$, and $\sigma_{xz}^x$. We attribute the enormity of this response to large extended regions in the Brillouin zone (shaded in Fig. \ref{fig:band-SHC}) where the band structure is nearly degenerate. 
We also determined that the bulk spin Hall conductivity is very robust and stable to a number of symmetry breaking terms, provided that the aforementioned nearly degenerate band structure is preserved. Our thin film calculations implicated the modification of the bulk-like wave functions in the thin film to be responsible for enhanced SHE in certain geometries.
This thin film consideration is unlike the symmetry-broken-augmented bulk calculations where the symmetry breaking perturbations do not induce as large a change to the electronic states. 
The nature of the electronic states being a key ingredient in the spin Hall conductivity suggests that further constrained and restricted geometries of SrIrO$_3$ (where the electronic states can change significantly due to, for example, certain symmetries being decisively broken) can lead to large enhancement of the spin Hall conductivity with respect to the corresponding bulk response. Although this study examines the case of SrIrO$_3$, we are hopeful that similar 5d-like iridate materials can also have large spin Hall conductivities. This aspiration is partly rooted in recent experimental work in binary 5d transition metal oxide IrO$_2$, where large SHE was demonstrated with high spin current conversion.\cite{Fujiwara_2013} 

{The results from our theoretical calculations seem to agree at least qualitatively with the recent experimental report where large spin Hall conductivity was observed in SrIrO$_3$ thin films.} To enable a more quantitative comparison with experimental spin Hall conductivity measurements, it is important to take into account the effects of the neighbouring ferromagnetic permalloy layer (Py) on the SHC. Although we provide some early analysis of these effects in Appendix \ref{app_mag_perturbations}, we propose taking into account the effects of the Py exchange field on the interface to be an interesting direction for future research.

\section*{Acknowledgements}

K. H. and Y. B. K. thank T. Nan, T. J. Anderson, C. B. Eom, M. S. Rzchowski, J. Gibbons, and D. C. Ralph for helpful discussions and collaborations in a related study. This work was supported by the NSERC of Canada and the Center for Quantum Materials at the University of Toronto. We acknowledge the hospitality at the Kavli Institute for Theoretical Physics, supported in part by the NSF grant PHY-1125915 and the Aspen Center for Physics, supported in part by NSF grant PHY-1607611, where some parts of this work were done. H.W.L was supported by the National Research Foundation of Korea (NRF) grant (No. 2011-0030046). Some of the computations were performed on the GPC supercomputer at the SciNet HPC Consortium.\cite{scinet} SciNet is funded by: the Canada Foundation for Innovation under the auspices of Compute Canada; the Government of Ontario; Ontario Research Fund - Research Excellence; and the University of Toronto.

\appendix

\section{Hamiltonian matrix $H_{\bf k}$} \label{app_H}

The explicit form of the tight binding model \cite{Carter_2012,Chen_2015} is given by the 8$\times$8 Hamiltonian matrix:
\begin{equation}
\begin{aligned}
H_{\bf {k}} = \ & \left(\epsilon_{r, \bf{k}} ^{po} \sigma_y + \epsilon_{i, \bf{k}} ^{po} \sigma_x \right) \nu_z \tau_y + \left(\epsilon_{r, \bf{k}} ^{zo} \sigma_y + \epsilon_{i, \bf{k}} ^{zo} \sigma_x \right) \nu_y \tau_z \\
				     & + \left(\epsilon_{r, \bf{k}} ^{do} \sigma_y + \epsilon_{i, \bf{k}} ^{do} \sigma_x \right) \nu_x \tau_y + \epsilon_{r, \bf{k}} ^d \nu_x \tau_x + \epsilon_{i, \bf{k}} ^d \nu_y \tau_y \\
				     & + \epsilon_{r, \bf{k}}^p \tau_x + \epsilon_{i, \bf{k}}^p \sigma_z \tau_y + \epsilon_{\bf{k}} ^z \nu_x  + \lambda_{\bf{k}} .
\end{aligned}
\end{equation}
Here the Pauli matrices $\sigma, \tau, \nu$ act on the spin ($\uparrow, \downarrow$) and sublattice (1,2,3,4) degrees of freedom in the following way.
\begin{equation}
\begin{aligned}
\sigma_x: & ~ \uparrow ~ \leftrightarrow ~ \downarrow,
\\
\nu_x: & ~ 1 ~ \leftrightarrow ~ 3, ~ 2 ~ \leftrightarrow ~ 4,
\\
\tau_x: & ~ 1 ~ \leftrightarrow ~ 2, ~ 3 ~ \leftrightarrow ~ 4 .
\end{aligned}
\end{equation}
The coefficients of $H_{\bf k}$ are:
\begin{equation}
\begin{aligned}
&\lambda_{\bf{k}} = t_{xy} \cos{k_x} \cos{k_y},  \\
&\epsilon_{r, \bf{k}} ^{p} = 2t_p\left(\cos{k_x} + \cos{k_y} \right), \\
&\epsilon_{i, \bf{k}} ^p = - t_{p} ' \left(\cos{k_x} + \cos{k_y} \right), \\
&\epsilon_{\bf{k}} ^z = 2t_p \cos{k_z}, \\
&\epsilon_{r, \bf{k}} ^{zo} = t_z ^o \cos{k_z}, \\
&\epsilon_{i, \bf{k}} ^{zo} = - t_z ^o \cos{k_z}, \\
&\epsilon_{r, \bf{k}} ^{d} = t_d \left(\cos{k_x} + \cos{k_y} \right) \cos{k_z},  \\
&\epsilon_{i, \bf{k}} ^{d} = t_d '  \left(\sin{k_x} + \sin{k_y} \right) \sin{k_z},  \\
&\epsilon_{r, \bf{k}} ^{po} = t_{1p} ^o \cos{k_x} + t_{2p} ^o \cos{k_y}, \\
&\epsilon_{i, \bf{k}} ^{po} = - t_{2p} ^o \cos{k_x} - t_{1p} ^o \cos{k_y}, \\
&\epsilon_{r, \bf{k}} ^{do} = t_d ^o \sin{k_y} \sin{k_z}, \\
&\epsilon_{i, \bf{k}} ^{do} = t_d ^o \sin{k_x} \sin{k_z},
\end{aligned}
\end{equation}
where $k_x\equiv{\bf k}\cdot{\bf x}$, $k_y\equiv{\bf k}\cdot{\bf y}$, $k_z\equiv{\bf k}\cdot{\bf z}$, and $\{ t_p, t_z, t_{xy}, t_d, t'_d, t'_p, t_{1p}^o, t_{2p}^o, t_{z}^o, t_{d}^o \}$ are hopping parameters.
The values of the hopping parameters used in our calculations are listed in Table \ref{tab:hopping-para}. 
\begin{table}[h]
\caption{Hopping parameters of the tight-bind model in the unit of eV.\cite{Carter_2012,Chen_2015}}
\begin{ruledtabular}
\begin{tabular}{cccccccccc}
$t_p$ & $t_p^{'}$ & $t_{xy}$ & $t_z$ & $t_z^{0}$ & $t_d$ & $t_d^{'}$ & $t_{1p}^0$ & $t_{2p}^0$ & $t_d ^{0}$
\\
\hline
-0.6 & -0.15 & -0.3 & -0.6 & 0.13 & -0.3 & 0.03 & 0.1 & 0.3 & 0.06
\\
\end{tabular}
\end{ruledtabular}
\label{tab:hopping-para}
\end{table}

\section{Representations of Symmetry Operators} \label{Perturb_op}

{
Here we provide representations of other symmetry operators: inversion ($\bar{I}$), time-reversal ($T$), and mirror reflection ($m$).
}
\begin{equation}
\begin{array}{ccl}
\bar{I}: \psi_{\bf k} &\rightarrow&  \psi_{(-k_a,-k_b,-k_c)},
\\
\\
T: \psi_{\bf k} &\rightarrow& -i\sigma_y \mathcal{K} \psi_{(-k_a,-k_b,-k_c)},
\\
\\
m: \psi_{\bf k} &\rightarrow& -i e^{-ik_c/2} \sigma_z \nu_x \psi_{(k_a,k_b,-k_c)},
\end{array}
\label{eq:symm-transformation-2}
\end{equation}
{
where $\mathcal{K}$ means complex conjugation.
In the representations of the above equation and Eq. \ref{eq:symm-transformation},
phase factors can vary depending on the gauge choice for the electron operators $\psi_{\bf k}$.
Despite such gauge dependence, phase factors contain important physical information such as the commutation relations in Eq. \ref{eq:commutation}.
}
\newpage
\section{Influence of perturbations on bulk band structure} \label{app_perturbations}

\begin{figure}[h]
\centering
\subfloat[UR Direction. \label{fig:gappedUR}]{%
  \centering
  \includegraphics[width=.5\linewidth]{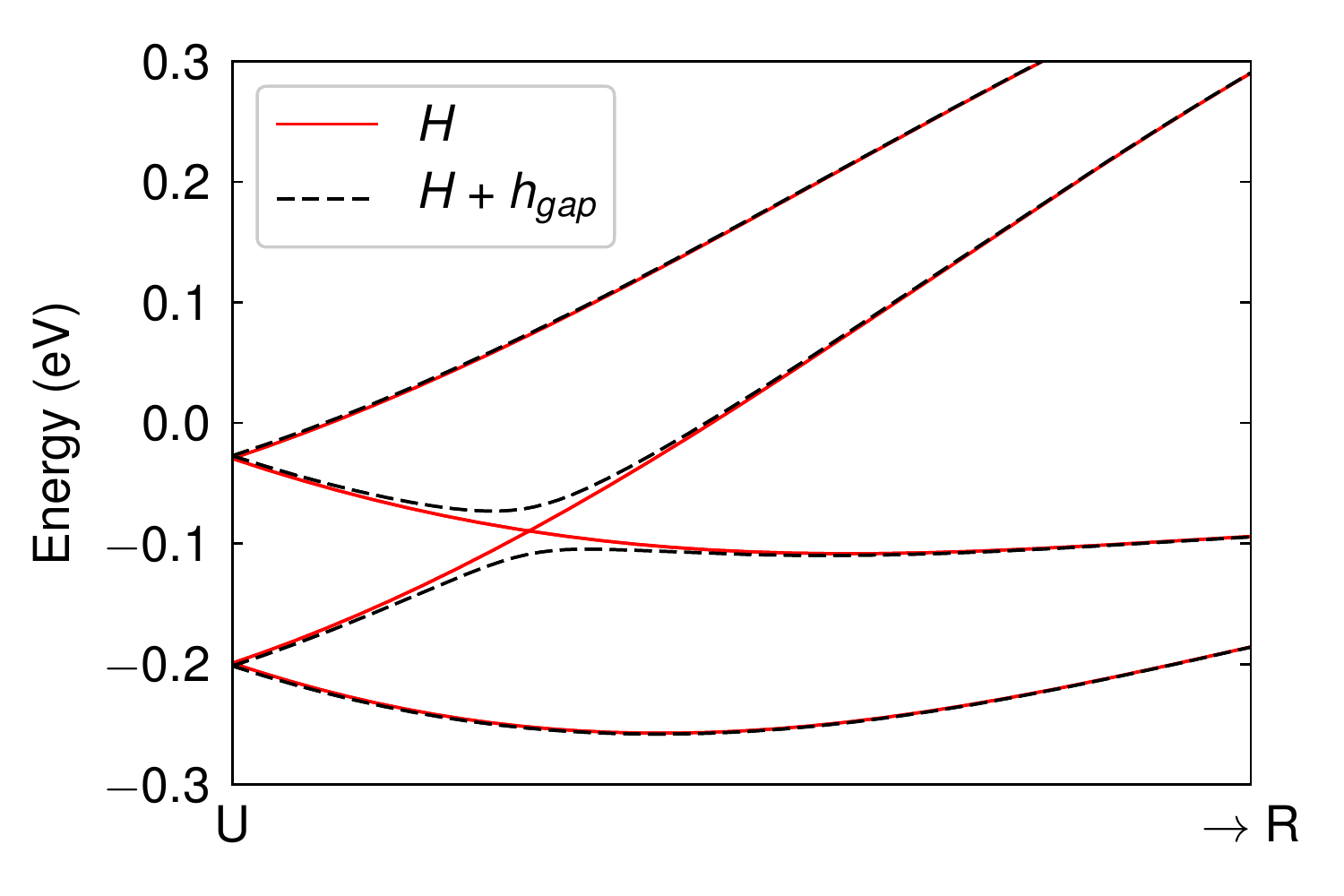} 
 }
\subfloat[UX Direction. \label{fig:gappedUX}]{
  \centering
  \includegraphics[width=.5\linewidth]{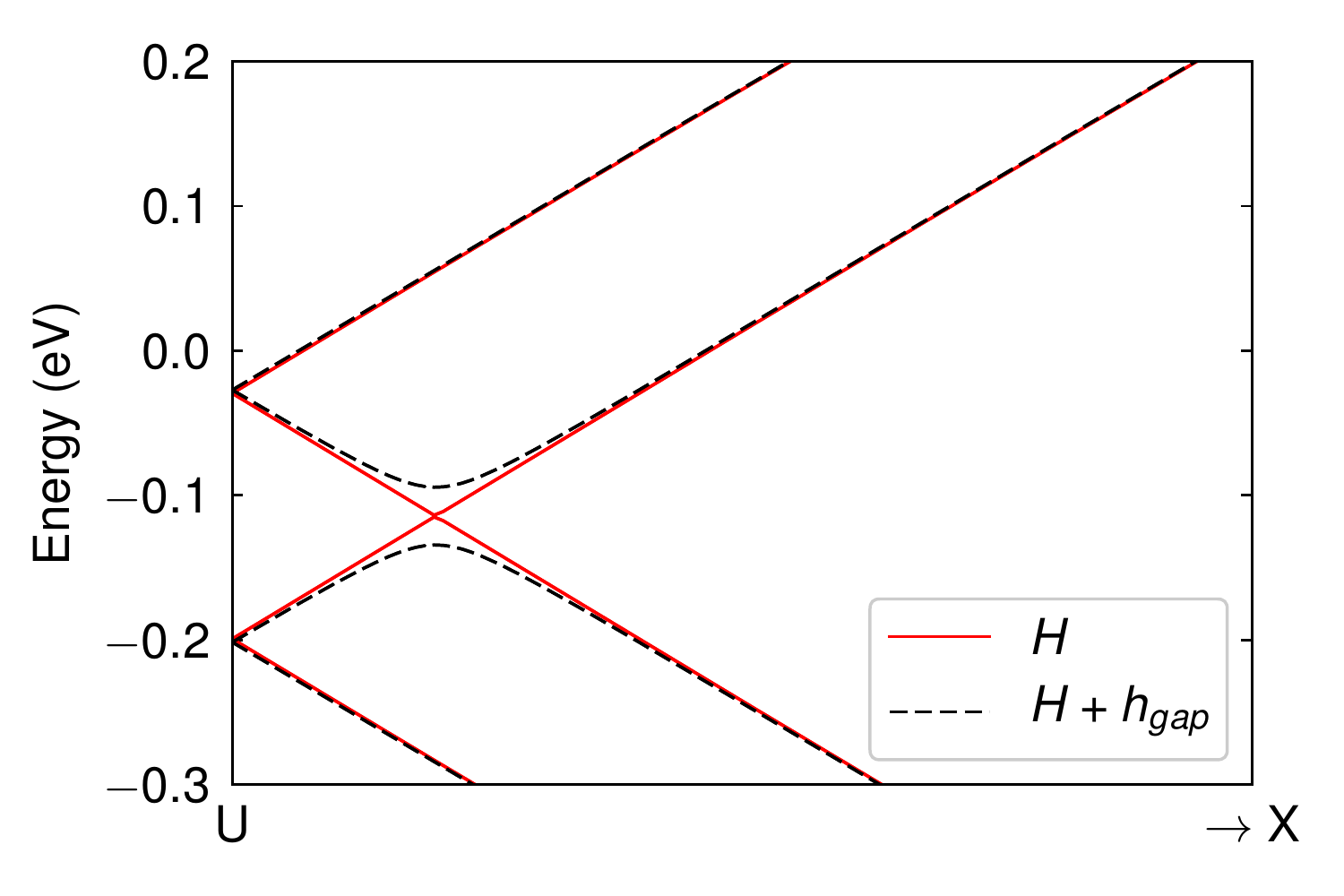} 
 }
\caption{Gapped nodal ring band structure due to $h_{gap}$ with $t_{gap}=0.01$eV.}
\label{fig:gapped_together}
\end{figure}

\begin{figure}[h]
\centering
\subfloat[UR Direction. \label{fig:xxUR}]{%
  \centering
  \includegraphics[width=0.5\linewidth]{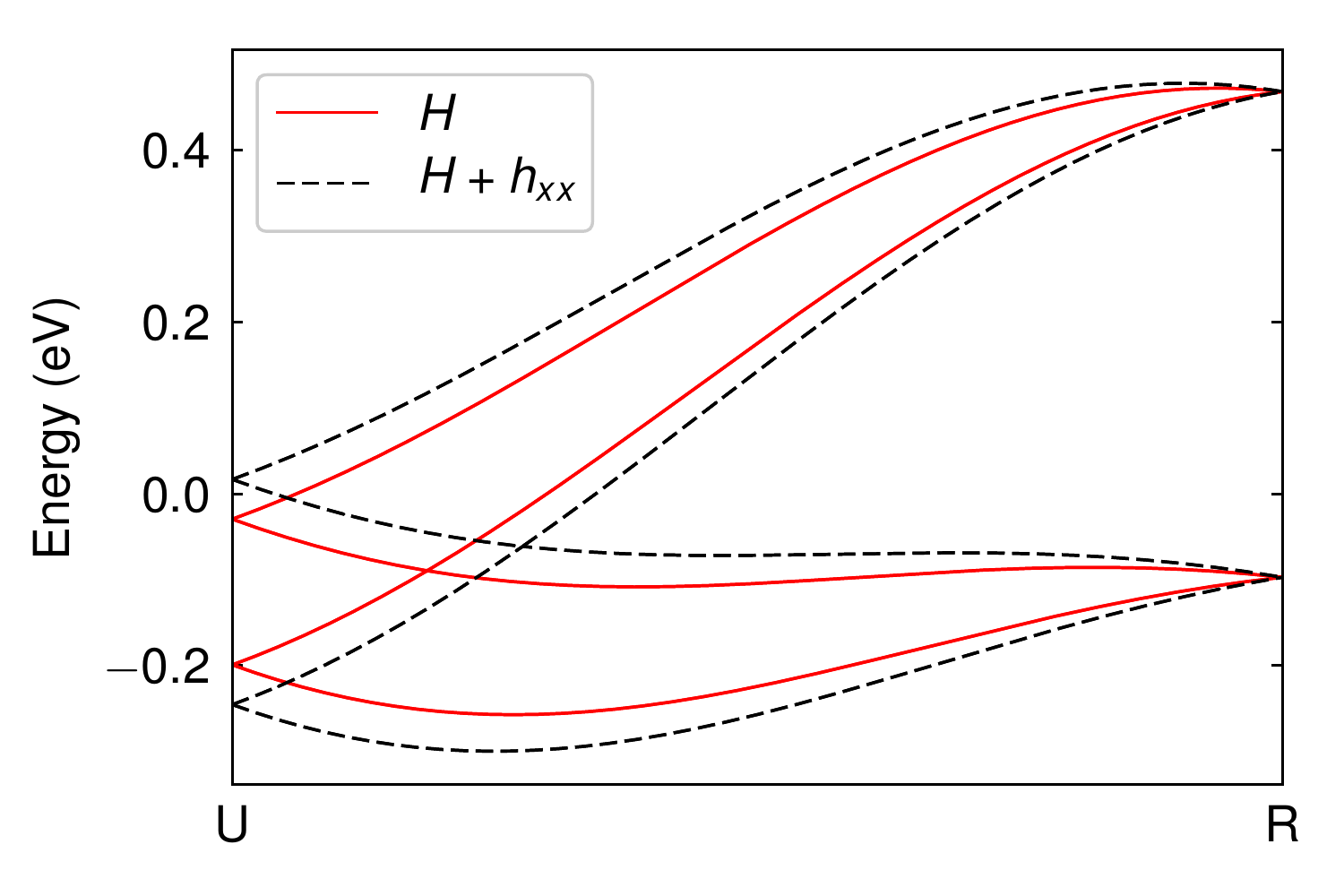} 
}
\subfloat[UX Direction. \label{fig:xxUR}]{%
  \centering
  \includegraphics[width=0.5\linewidth]{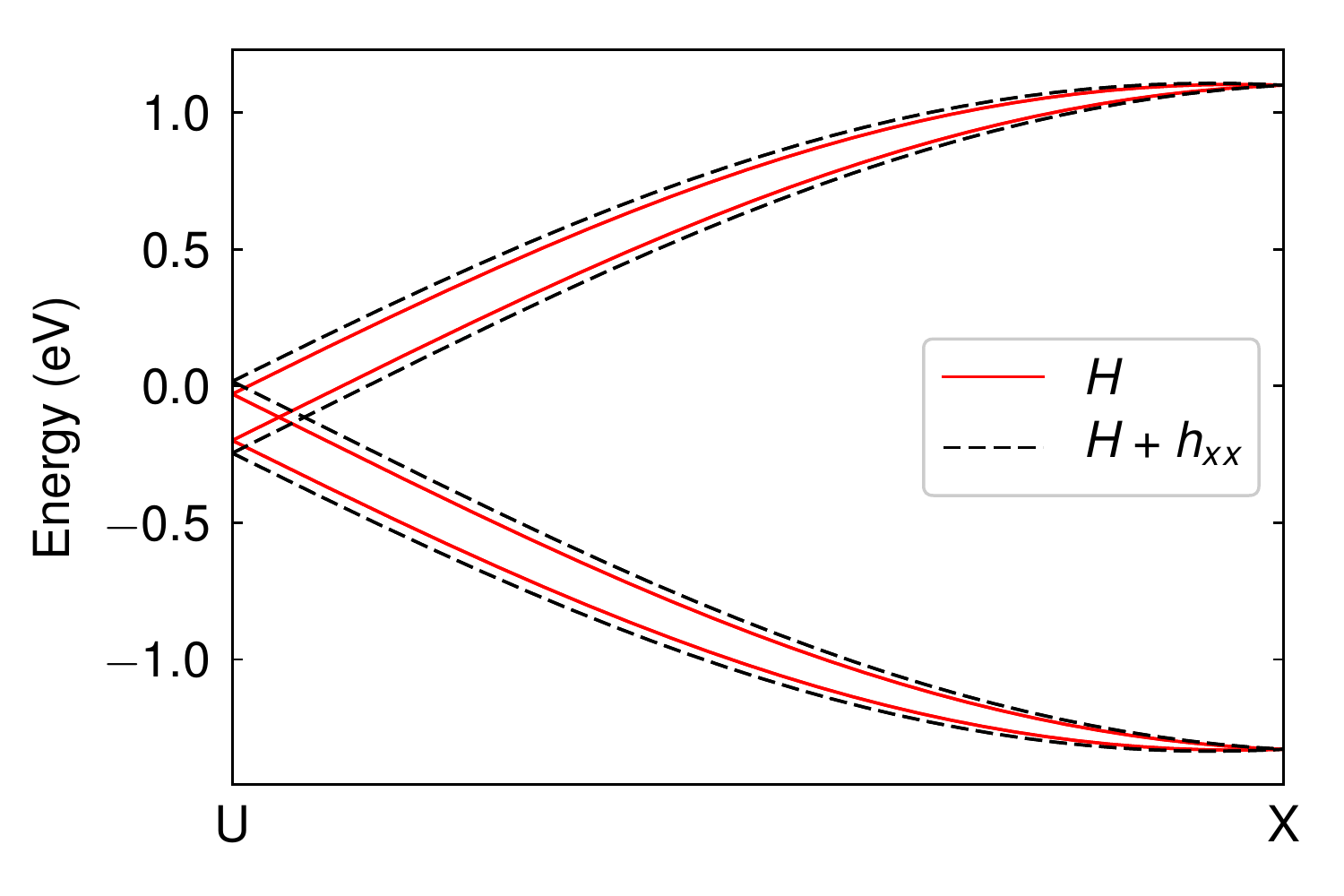}
}
\caption{Perturbed band structure due to $h_{xx}$ with $t_{xx}=0.05$eV.}
\label{fig:xx_together}
\end{figure}

\begin{figure}[!h]
\centering
\subfloat[TS Direction. \label{fig:xzTS}]{%
  \centering
  \includegraphics[width=0.5\linewidth]{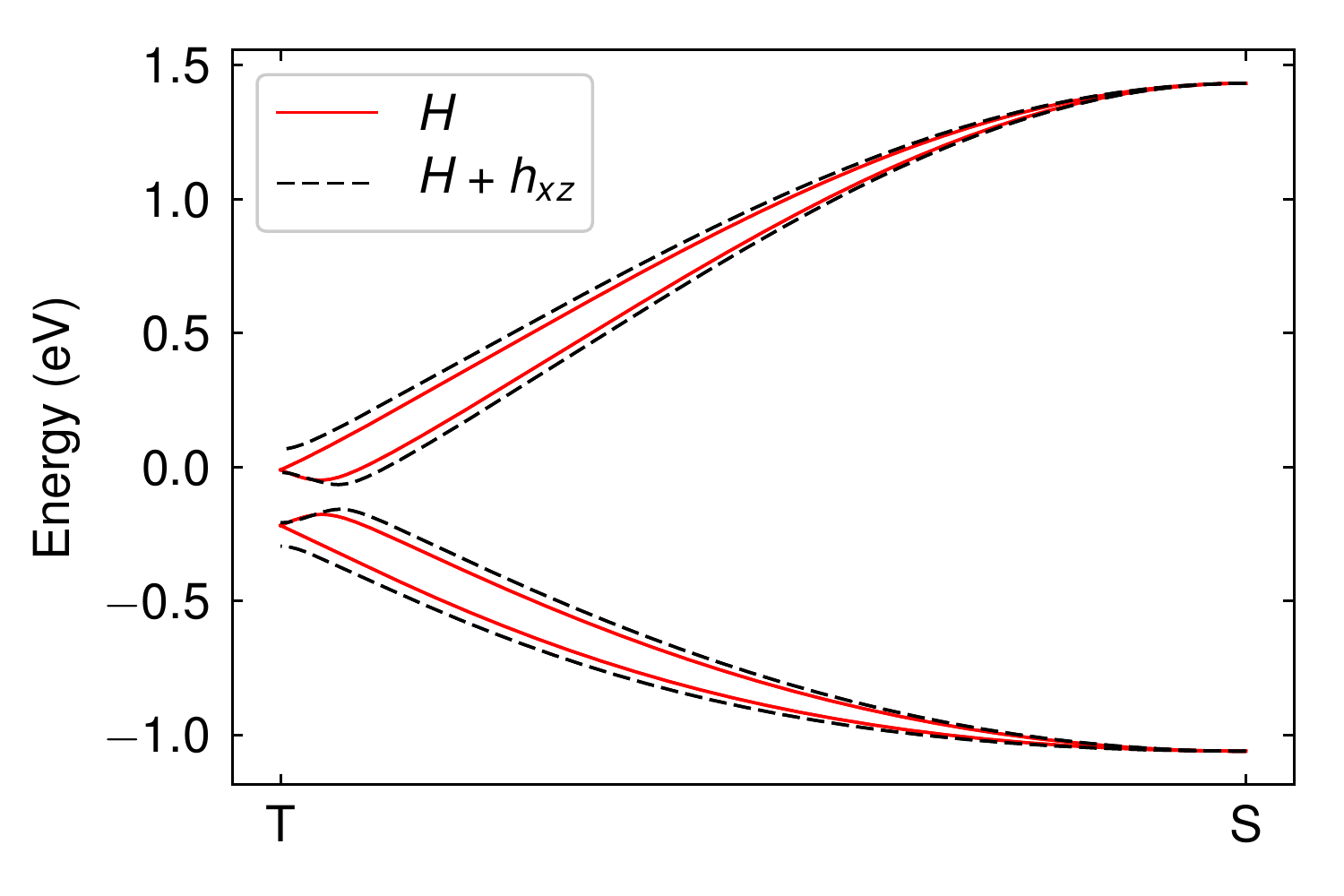} 
}
\subfloat[TY Direction. \label{fig:xzTY}]{%
  \centering
  \includegraphics[width=0.5\linewidth]{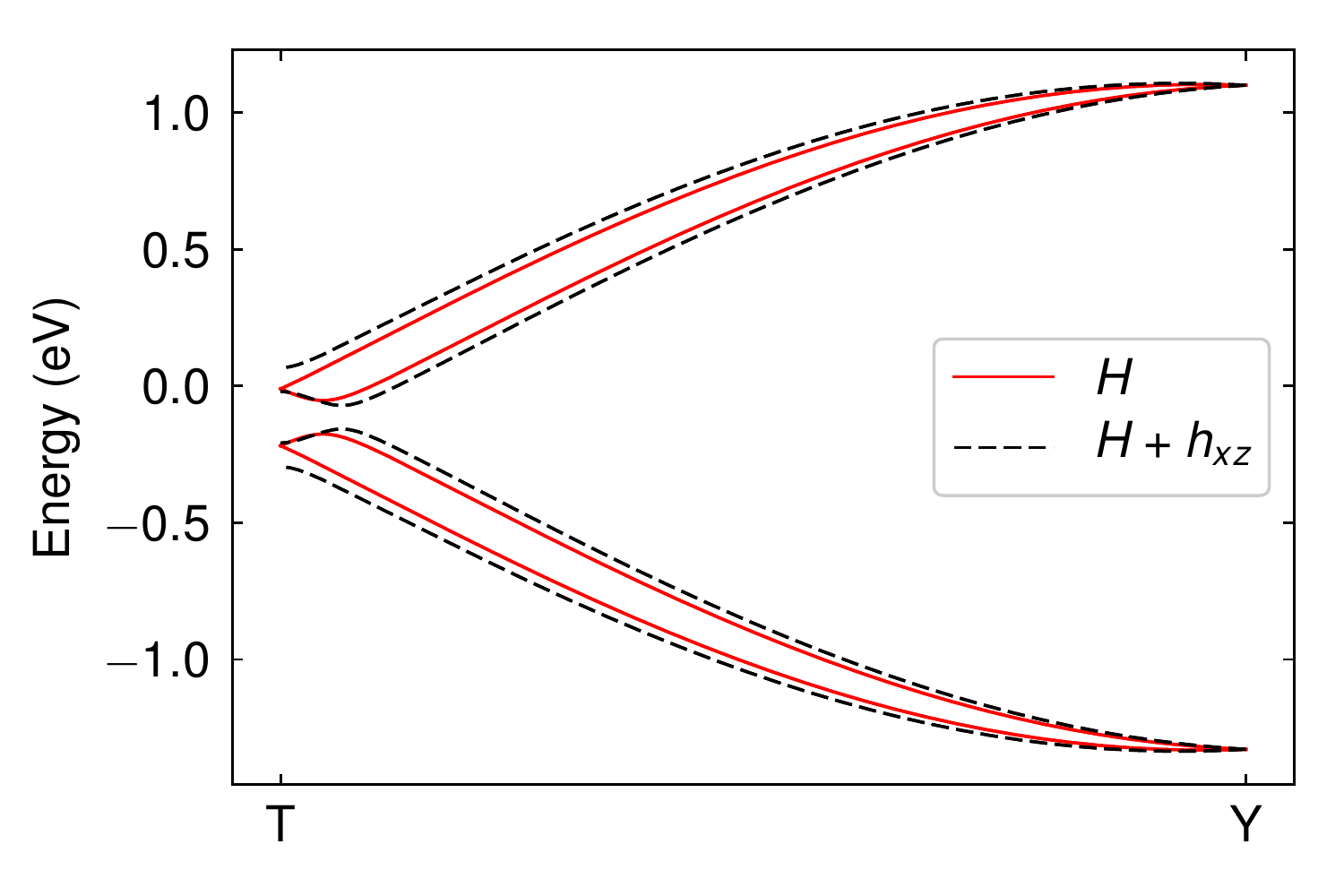}
}
\caption{Perturbed band structure due to $h_{xz}$ with $t_{xz}=0.05$eV.}
\label{fig:xz_together}
\end{figure}

\section{Influence of magnetic perturbations on bulk spin Hall effect} \label{app_mag_perturbations}

\begin{figure}[h]
\centering
\includegraphics[width=0.9\linewidth,angle=0]{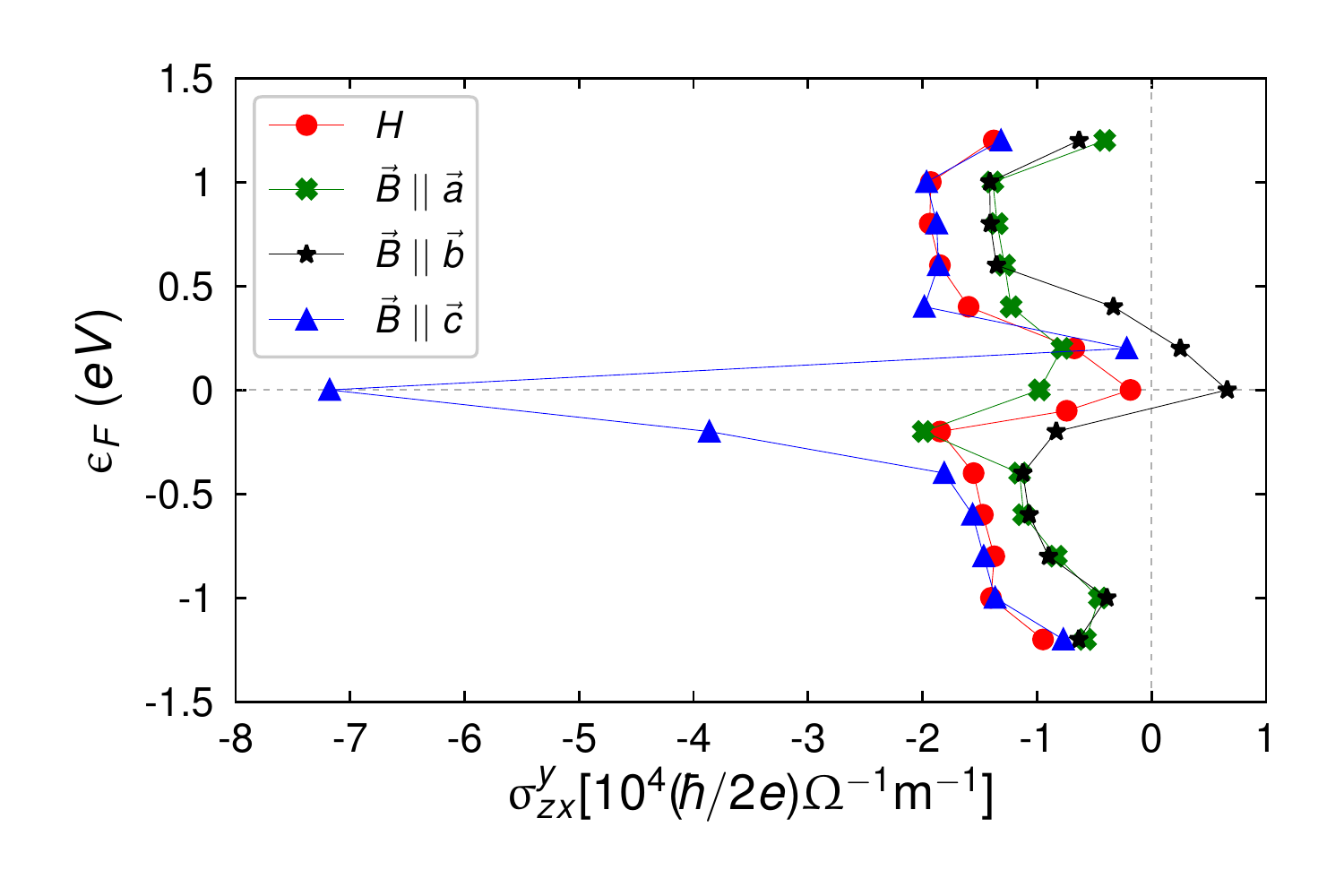}
\caption{SHC of original bulk model (${H}$) compared to bulk model augmented by magnetic (Zeeman) perturbations.}
\label{fig:SHC_mag}
\end{figure}

In the main text, we considered the influence of non-magnetic perturbations to determine the stability of the spin Hall conductivity. However as discussed in the Sec. \ref{conclusion}, a ferromagnetic permalloy, Py,  (situated nearby to the SrIrO$_3$ thin film) has an exchange field that could potentially leak into the SrIrO$_3$. The Py plays an important role as it is used in the actual measurement of the spin Hall effect: the spin current which flows into the Py (from the spin Hall effect in SrIrO$_3$) exerts a torque on the magnetic moment causing it to precess. The precession leads to a voltage being generated (via the anisotropic magnetoresistance) in the Py which is measured. Our goal here is to incorporate the effects of the Py's exchange field into our calculations. In order to get an estimate for the strength of the ferromagnetic exchange field, one notes that the exchange stiffness constant in permalloy to be $\mathcal{A}_{exchange} = 1.3 \times 10^{-11} J/m$ \cite{permalloy} and the lattice constant to be $a = 3.55$ {\AA}.\cite{permalloy_thick} Hence, an estimate of the exchange constant is  $J_{exchange} \approx \mathcal{A} a = 0.0288 eV$. This corresponds to an exchange field of approximately $B_{ex} \approx 490T$. It is important to note that since the actual exchange field is localized at the interface, as a simple approximation we simulate its impact as a constant Zeeman field that acts uniformly on the bulk. Moreover, we take the strength of this Zeeman field to be much weaker than the actual exchange field to incorporate the fact that the exchange field is localized at the interface and gets weaker for distances further away from the interface: $B_{Zeeman} = 0.01$ eV $\leftrightarrow \approx 170$T. This Zeeman magnetic field perturbation also serves as a means to determine the robustness of the spin Hall effect in SrIrO$_3$ to large, stray external magnetic fields. Since the direction of the exchange field from the permalloy can be arbitrary, we examine the effects of the Zeeman field pointing along each one of the orthorhombic axes: $\{ \vec{a}, \vec{b}, \vec{c} \}$. As has been previously described,\cite{Chen_2015} the nodal ring undergoes different experiences depending on the direction of the external magnetic field.  If the magnetic field is along the $\vec{c}$ axis, the chiral symmetry is broken and the nodal ring is gapped out (in addition to the two-fold degeneracy being broken due to broken time-reversal symmetry). If the magnetic field is along the $\vec{b}$ axis, the doubly degenerate nodal ring splits into two non-degenerate nodal rings shifted along the $\vec{c}$ axis. Finally, if the magnetic field is along the $\vec{a}$ axis, the nodal ring is replaced by 3D Dirac nodes (which are not symmetry protected). Below we present the impact of the nodal ring calculation on the $\sigma_{zx}^y$ component in Fig. \ref{fig:SHC_mag}.

\begin{figure}[h]
\centering
\includegraphics[width=0.8\linewidth,angle=0]{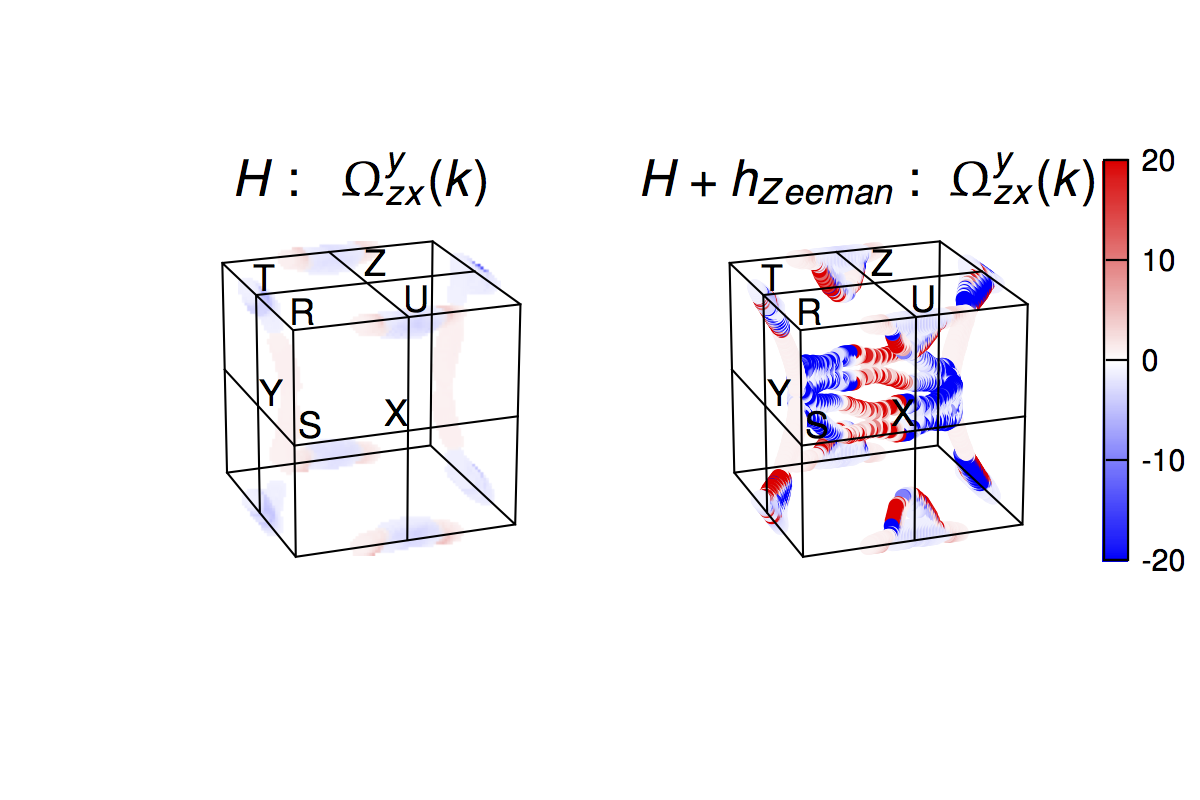}
\caption{Momentum-resolved SHC at the zero Fermi level for $\sigma_{zx}^y$ for original bulk model (${H}$) compared to bulk model augmented by magnetic (Zeeman) perturbations.}
\label{fig:berry_mag}
\end{figure}

The spin Hall conductivity is stable to the introduction of magnetic fields along the $\vec{a}$ and $\vec{b}$ axes. A magnetic field along the $\vec{c}$ axis does not change the spin Hall conductivity substantially for all Fermi levels except near the zero energy level. In particular, at the zero energy level, $\sigma_{zx}^y$ increases by about an order of magnitude. To discern the cause for this, we once again examine the momentum-resolved SHC at the zero Fermi level in Fig. \ref{fig:berry_mag}. There appear to be new features that develop throughout the Brillouin zone (rings around the $U$ and $T$ points, as well as a circular loop about the zone centre) that lead this large change in the spin Hall conductivity. Although this increase is very promising, one should recall again that in a realistic system the exchange field is localized near the boundary. Nevertheless, one can still appreciate that the spin Hall conductivity is stable to such large magnetic fields for most of the Fermi levels (except near the zero energy level). Near the zero energy level, these results suggest that the exchange field can indeed induce a large change in the spin Hall conductivity.

\newpage
\bibliography{references_bibtex}

\end{document}